\documentclass[10pt,preprint]{aastex}
\shorttitle{GRB Progenitors}
\shortauthors{Fryer et al.}

\begin{document}
\title{Constraints on Type Ib/c and GRB Progenitors}

\author{Chris L. Fryer\altaffilmark{1,2}, Paolo
A. Mazzali\altaffilmark{3,4}, Jason Prochaska\altaffilmark{5}, Enrico
Cappellaro\altaffilmark{6}, Alin Panaitescu\altaffilmark{7}, Edo
Berger\altaffilmark{8,9,10}, Maurice van Putten\altaffilmark{11}, Ed P.J.
van den Heuvel\altaffilmark{12}, Patrick Young\altaffilmark{13,14},
Aimee Hungerford\altaffilmark{2}, Gabriel Rockefeller\altaffilmark{2},
Sung-Chul Yoon\altaffilmark{12}, Philipp
Podsiadlowski\altaffilmark{15}, Ken'ichi Nomoto\altaffilmark{16},
Roger Chevalier\altaffilmark{17}, Brian Schmidt\altaffilmark{18}, Shri
Kulkarni\altaffilmark{19}}

\altaffiltext{1}{Department of Physics, The University of Arizona,
Tucson, AZ 85721}
\altaffiltext{2}{CCS Division, LANL, Los Alamos, NM 87545}
\altaffiltext{3}{INAF-OATs, via Tiepolo, 11, 34131, Trieste, Italy}
\altaffiltext{4}{Max-Planck-Institut f\"ur Astrophysik,
Karl-Schwarzschild-Str. 1, 85741 Garching bei M\"unchen, Germany}
\altaffiltext{5}{University of California Observatories-Lick Observatory, 
University of California, Santa Cruz, CA 95064}
\altaffiltext{6}{INAF-Osservatorio Astronomico, vicolo
dell'Osservatorio 5, 35122 Padova, Italy}
\altaffiltext{7}{ISR Division, LANL, Los Alamos, NM 87545}
\altaffiltext{8}{Observatories of the Carnegie Institution of
Washington, 813 Santa Barbara Street, Pasadena, CA 91101}
\altaffiltext{9}{Princeton University Observatory, Peyton Hall, Ivy
Lane, Princeton, NJ 08544}
\altaffiltext{10}{Hubble Fellow}
\altaffiltext{11}{LIGO Laboratory, MIT 17161, Cambridge, MA 02139, USA}
\altaffiltext{12}{Astronomical Institute Anton Pannekoek, University
of Amsterdam, Kruislaan 403, 1098 SJ, Amsterdam, The Netherlands}
\altaffiltext{13}{Astronomy Department, The University of Arizona,
Tucson, AZ 85721}
\altaffiltext{14}{X Division, LANL, Los Alamos, NM 87545}
\altaffiltext{15}{Department of Astrophysics, University of Oxford,
Denys Wilkinson Building, Keble Road, Oxford OX1 3RH, UK}
\altaffiltext{16}{Department of Astronomy and RESCEU, University of
Tokyo, Hongo 7-3-1, Bunkyo-ku, Tokyo 113-0033, Japan}
\altaffiltext{17}{Department of Astronomy, University of Virginia,
P.O. Box 400325, Charlottesville, VA 22904-4325}
\altaffiltext{18}{Mount Stromlo and Siding Spring Observatory, Private
Bag, Weston Creek P.O., Canberra, ACT 2601, Australia}
\altaffiltext{19}{Caltech Optical Observatories, Mail Code 105-24,
California Institute of Technology, 1200 East California Boulevard,
Pasadena, CA 91125}

\begin{abstract}
  
  Although there is strong support for the collapsar engine as the
  power source of long-duration gamma-ray bursts (GRBs), we still do
  not definitively know the progenitor of these explosions.  Here we
  review the current set of progenitor scenarios for long-duration
  GRBs and the observational constraints on these scenarios.
  Examining these, we find that single-star models cannot be the only
  progenitor for long-duration GRBs.  Several binary progenitors can
  match the solid observational constraints and also have the
  potential to match the trends we are currently seeing in the
  observations.  Type Ib/c supernovae are also likely to be produced
  primarily in binaries; we discuss the relationship between the
  progenitors of these explosions and those of the long-duration GRBs.

\end{abstract}


\section{Introduction}

As early as 185 AD, the energetic transients known as supernovae have
excited the imagination of mankind (Stephenson \& Clark 1976, Chin \&
Huang 1994).  We now believe (and even know for a fact in some cases)
that many of these supernovae (types Ib, Ic, and II) arise from the
collapse of massive stars.  Theorists have gradually converged on a
mechanism that takes the potential energy released in the collapse of
a stellar core down to a neutron star and injects energy into the
convective region above the neutron star, driving an explosion (see
Fryer 2003 for a review).  However, the details (including the
relevant physics) of this explosion mechanism are far from settled. It
is important to remember that ``supernova'' has a phenomenological
definition. Any event that disrupts a star with sufficient violence
will be observed as a supernovae. One of the biggest uncertainties in
determining the explosion mechanism is our understanding, or lack
thereof, of the supernova progenitor.  Although it is likely that the
explosion arises from the collapse of a massive ($\gtrsim
8$\,M$_\odot$) star, the exact nature of the evolution of this
progenitor is unknown, especially for Type Ib/c supernovae.

One of the latest developments in the study of explosions from
collapsing massive stars has been the discovery of gamma-ray bursts
(GRBs).  Engines invoking the collapse of massive stars have once
again become the favored mechanisms behind a class (the long-soft
burst class) of GRBs.  Observations of these long-duration bursts, such
as the association of GRBs with star-forming galaxies and
star-formation regions in galaxies (Fruchter et al. 2006), have added
support to this model.  But the most convincing observational evidence
has been the concurrent and cospatial ``supernova-like'' outbursts
associated with GRBs.  These ``supernova-like'' bursts are evidence
that the GRB explosion is part of the disruption of a massive star.

This association between GRBs and ``supernova-like'' explosions has
led to the appearance of a new class of stellar explosion, the
so-called ``hypernova''.  A number of definitions for hypernovae
exist, from the energetic outburst produced by a collapsar (Woosley
1993; Paczynski 1998) to the supernova associated with GRB outbursts
\footnote{For a while, Woosley wanted to use the term hypernova to
rename pair-instability supernovae.}.  Our definition is a bit
broader; we use the term hypernova to denote all explosions that
exhibit stronger-than-normal (more than a few times $10^{51}$\,erg) explosion
energies and/or with strong evidence for asymmetries (Nomoto et al. 2005)
\footnote{Note that some hypernovae can have relatively normal
explosion energies, but are still classified as hypernova based on the
large asymmetries: e.g. 2006aj (Soderberg et al. 2006).}.  With this
definition, the ``supernovae'' associated with GRBs are a subset of
the hypernova class.

The relationship between normal supernovae and hypernovae has led to
intense discussion with views ranging from ``all supernovae are
hypernovae and the current models of supernovae are all wrong'' to
``hypernovae have nothing to do with the explosions of massive
stars''.  The former interpretation essentially ignored the bulk of
the existing supernova observations and has finally been put to rest
in the GRB community by comparisons between supernovae and hypernovae
(Soderberg et al. 2006).  The latter interpretation seems unlikely
given how well massive star models fit the observed hypernovae (e.g.
Deng et al. 2005; Mazzali et al. 2006; Maeda et al. 2006).  We take a
more moderate interpretation, assuming that hypernovae are a rare set
of massive star explosions with an engine different from the standard
supernova model.

Although the evidence suggesting that these explosions are produced
inside massive stars continues to grow, we know very little else about
the engine behind hypernovae and GRBs.  Based primarily on the fact
that these explosions are different from ``normal'' supernovae,
theorists have argued that the engine itself must also be different.
The leading theory, the collapsar engine (Woosley 1993), suggests that
the explosion is produced after the massive star collapses to a black
hole.  The energy released during accretion of the infalling stellar
material onto this black hole provides the energy for the explosion
(via neutrino annihilation or a magnetic field transfer mechanism)
(Narayan et al. 1992).  But this can only occur if the energy does not
also accrete onto the black hole.  The requirement for GRBs, then, is
that the infalling material have sufficient angular momentum to hang
up in a disk before accreting.

Unfortunately, hypernovae are rare events (roughly 1000 times less
frequent than normal supernovae), and their rarity allows theorists
the freedom to devise all manner of exotic formation scenarios for the
progenitors of these explosions.  In this paper, we review the wide
range of possible progenitors and try to constrain them with the
current set of observational data.  One clue may be that, so far, all
hypernovae (with and without GRB jets) have been classified as type
Ib/c supernovae; i.e.\ these bursts do not have strong hydrogen lines
in their spectra.  Indeed, hypernovae do not even have strong helium
lines in their spectra, suggesting that the progenitors of these
explosions have lost much of their helium layer (we will discuss this
in more detail in \S 3).  It may well be that the progenitors of
hypernovae are merely a subset of the progenitors of their more common
cousins, type Ib/c supernovae.  We will review these progenitors as
well to better understand the link between supernovae and gamma-ray
bursts.

In this paper, we focus our studies of GRBs on progenitors of the
collapsar engine, but we include variants often neglected in
discussions of collapsar progenitors.  A wide variety of progenitors
has been proposed, each including a set of predictions for
characteristics that presumably can be compared to observations.  We
discuss the progenitors and their characteristics in \S 2.  Studies of
the observed hypernova samples (with and without GRB jets) have
produced a number of constraints that have been used to argue for or
against certain progenitors.  However, it is quite difficult to make
definitive observational statements on the current set of progenitor
predictions.  In \S 3, we review the current state of observational
constraints.  The goal of this paper is to draw the attention of
theorists to the firm observational constraints and of observers to
the firm theoretical predictions to provide a road map for the future
that will increase the amount of information in the intersection of
these two data sets.  We conclude with a review of how current models
fare with the existing data.

\section{Progenitors}

\subsection{Progenitors for Ib/c supernovae}

One possible picture for the origin of type Ib/c supernovae is that
the progenitors for these explosions are the most massive stars (see
Hirschi et al. 2004 and references therein).  These stars have very
strong winds, which ultimately cause the stars to lose their entire
hydrogen envelope and become strong Wolf-Rayet stars.  However, when
Heger et al. (2003) studied single stars, they found that very few
non-rotating single stars at solar metallicity eject their entire
hydrogen envelopes in winds.  Figure 1 shows the fraction of
collapsing stars that form type II and type Ib/c supernovae as a
function of metallicity from this Heger et al. (2003) study.  The thin
lines denote those supernovae that Heger et al.  (2003) believed would
only produce weak supernovae (based on the analysis of Fryer 1999).
Note that single non-rotating stars only produce Ib/c supernovae at
metallicities above 0.02$Z_\odot$.  But we expect these supernovae to
be have weak shocks and hence eject very little nickel.  Without the
high shock temperatures and the radioactive nickel to power the
emission and be, these supernovae will be dim.  Strong
Ib/c supernovae are not produced at all until the metallicity rises
above solar! This assumes, however, that the explosion is powered by
the standard neutrino-driven convection mechanism. An alternate engine
(such as a collapsar) may be able to produce bright supernovae from
these progenitors.

There are a few caveats to these results.  First, these results depend
sensitively on the mass-loss rates used and rates can shift along the
metallicity axis depending upon the values adopted for mass-loss.
However, if anything, the trend in the last decade has been that early
calculations have overestimated the mass loss.  Lowering this mass
loss would only push the minimum metallicity to form Ib/c supernovae
upward.  Note also that Yoon \& Langer (2005) have found that
rapidly-rotating stars can mix their hydrogen envelopes, effectively
removing the hydrogen envelope by burning it into helium.

Alternatively, these Ib/c supernovae could be formed in binaries
(Podsiadlowski, Joss, \& Hsu 1992).  Mass transfer in binaries can
eject matter, forming helium stars (Ib/c progenitors) or stars with
peculiar hydrogen envelopes (II pec or II linear progenitors).  The
list of supernovae which show evidence of a binary companion continues
to grow: SN 1987A (Podsiadlowski et al. 1990), SN 1993J (Podsiadlowski
et al. 1993; Nomoto et al. 1993; Woosley et al. 1994, Maund et
al. 2004), Cas A (Young et al. 2006), SN 2001ig (Ryder et al. 2006),
and possibly Puppis A (Winkler et al. 1989) and SN 1994I (Sauer et
al. 2006).  Given the current uncertainties in stellar evolution, it
is difficult to really prove that an observed supernova came from a
binary system.  With the latest results showing that $\gtrsim 75$\%
(this value could be 100\%) of all massive stars are in close binaries
that will undergo mass transfer (Kobulnicky et al. 2006), type Ib/c
are primarily formed in binaries, unless supernova do not form from
massive stars.

Assuming that only one third of all stars are in close, interacting
binaries, Podsiadlowski et al. (1992) found that binaries would cause
roughly 15\% of all stellar collapses to form Ib/c supernovae.  With
the higher close-binary fraction estimated by Kobulnicky et al.
(2006), binaries would argue for a Type Ib/c supernova rate roughly
equal to 30\% of the total core-collapse supernova rate.  Because
there are multiple complementary channels that will produce Ib/c
supernovae, This estimate is not too dependent on binary mass
parameters (at least in the code used by Podsiadlowski et al. 1992 or
Fryer et al. 1998).  And although it depends upon the mass-loss rates
(and hence metallicity), the rate does not drop as dramatically with
decreasing metallicity as it does in single-star models (Fig. 2).  For
figure 2, we have used the population synthesis code of Fryer et al.
(1998), using their standard values for the population synthesis
parameters and varying only the mass-loss parameter.  The majority of
these type Ib/c supernovae are normal (i.e. not weak/dim) supernovae.

Figure 2 also shows the fraction of collapsing stars that lose not
only their hydrogen envelopes, but also $> 1$\,M$_\odot$ of their
helium envelopes.  This subset of all type Ib/c supernovae likely have
characteristics closer to type Ic supernovae.  Their rate is also
sensitive to mass-loss (i.e. metallicity); they account for 75\% of
all Ib/c supernovae at 10 times the Fryer et al. (1998) canonical
mass-loss coefficient but less than a third that total Ib/c rate at
lower mass-loss values.  For comparison, we also show the rate of
supernovae arising from progenitors that have lost 66\% of their
hydrogen envelopes.  These progenitors would produce peculiar or,
possibly, linear type II supernovae.

Both single-star and binary-star progenitor scenarios exist for type
Ib/c supernovae.  The two major differences in the predictions of
these formation scenarios are (1) their relative rates at low
metallicity and (2) the predicted number of weak to normal Ib/c
supernovae.  The single-star mixing model of Yoon \& Langer (2005) may
change the rate prediction for single stars, but it still predicts
that most Ib/c supernovae should produce weak supernova explosions
(recall that these weak supernovae are possible GRB progenitors).
Figure 3 shows the fate of massive stars (type Ib/c versus type II and
normal versus weak versus no supernova explosion from the standard
supernova mechanism) for the latest grid of mixing models from Yoon et
al. (2006).  Although this grid of stars, based on tables 4-7 in Yoon
et al. (2006), produces no normal type Ib/c supernovae, we must note
that the stellar evolution parameters may be tweaked to produce a
small amount of normal type Ib/c supernovae.  Obtaining a reliable
metallicity dependence of these ratios should easily distinguish these
progenitors.  But be aware that stellar evolution remains a field
involving many free parameters.  Although Heger et al. (2003) make
some solid claims to compare against observations, new features, such
as the mixing models of Yoon \& Langer (2005) can easily change these
predictions.  The predictions of binary models are, perhaps
surprisingly, a little more solid.  But since they too depend upon
stellar models, we must interpret any of these predictions with
caution as well.

\subsection{Progenitors for Hypernovae}
\label{hyper}

Most of our progenitors will focus on the collapsar engine with its three
basic requirements: i) the model must form a black hole in the center
of a star, ii) the model must produce sufficient angular momentum in
the star to form a disk around the black hole (but not too much
angular momentum to limit the accretion rate) and iii) the model must
eject the hydrogen envelope so that the jet produced by the collapsar
engine can punch out of the star.  We will broaden our scope to include any
progenitor that produces a non-degenerate star accreting rapidly onto
a compact remnant (either neutron star or black hole).  For most of
our progenitors, this opens up only a slightly broader range of
systems with rapid fallback.  It also includes such progenitors as the
He-merger scenario where a compact remnant spirals into the center of
its stellar companion (Fryer \& Woosley 1998).  The list of all the
progenitors studied and their basic predictions is given in
Table~\ref{table:progenitors}.  Before we discuss each progenitor
individually, let's first discuss the generic trends we expect from
these 3 constraints.

{\em Black Hole Formation:}

Nearly every GRB progenitor currently proposed requires the collapse
of a massive star down to a black hole (or a massive neutron star).
This constraint is equivalent to restricting GRB progenitors to those
stars that produce weak or no explosion under the standard
core-collapse supernova engine (see Fryer 2003 for a review).  It is
commonly assumed that the mass of the progenitor star determines
whether it will produce a strong or weak supernova explosion.
Although the exact mechanism behind core-collapse supernovae is not
known, current studies have focused on the role of the convective
engine between the surface of the neutron star and the accretion shock
of the infalling atmosphere.  If this convection region is indeed the
critical aspect of core-collapse determining the strength of the
standard supernova explosion, a consistent picture can be developed
describing the fate of a collapsing star as a function of its mass.
Under this assumption, Fryer (1999) argued that, in the absence of
stellar winds, the more massive stars ($\gtrsim 20$\,M$_\odot$) would
fail to produce strong explosions and collapse to form black holes.
His argument was based on two facts: i) the ram pressure at the top of
the convective region is larger for more-massive stars, making it more
difficult to explode these stars and leading to explosions that take
longer to develop and are weaker and ii) the binding energy of stellar
material increases dramatically with increasing star mass.  Fryer
argued that even though the uncertainties in the explosion engine were
great, these two combined effects allowed fairly accurate precision in
determining the transition between neutron star and black hole
formation ($23 \pm 5$M$_\odot$).

What Fryer (1999) had not considered were the uncertainties in stellar
evolution.  His results were entirely based on the Woosley \& Weaver
(1995) progenitors which did not include the effects of mass-loss from
stellar winds or rotation.  Fryer (2006) developed an analytic means
to estimate the final mass of the compact remnant after a supernova
explosion.  The results for several non-rotating pre-supernova
models are shown in Figure 4.  The lines show the results of the Woosley 
et al. (2002) progenitor models (dotted line refers to solar metallicity,
solid line refers to very low metallicity) and the points arise from
the Limongi \& Chieffi (2006) progenitor models (circle - solar, square - 0.2
solar, triangle - zero, metallicities).  For solar metallicity, all
massive stars produce weak explosions that then accrete through
fallback (above $\sim 20$\,M$_\odot$, these stars form black holes).

At lower metallicities, some stars will collapse directly down to
black holes, producing no explosion under the standard core-collapse
engine.  The original ``collapsar'' engine argued that GRBs are
produced by stars that collapse directly down to black holes (Woosley
1993).  If this is indeed a requirement, we find from the Heger et
al. (2003) progenitors that GRBs are not produced at solar
metallicity.  At low metallicity, stars above 30\,M$_\odot$ form GRBs
(solid line in Fig. 4).  The Limongi \& Chieffi (2006) progenitors
show similar trends at solar metallicity (no direct collapse GRBs at
solar metallicity), but exhibit quite different fates at zero
metallicity.  25\,M$_\odot$ stars at zero metallicity will collapse
directly to black holes.

Looking back at Figure 4, it is clear that above 20\,M$_\odot$, the
results of stellar evolution models vary drastically.  Figure 5 shows
the difference between the 25\,M$_\odot$ stars produced by Woosley et
al.  (2002) and by Limongi \& Chieffi (2006).  These differences are
believed to arise from different recipes for mass loss from stellar
winds and for convective mixing.  We will discuss mass loss below when
we discuss uncovering the hydrogen envelope.  As for mixing, recent
studies by Young et al. (2005) have shown that the structure of the
stellar core can change drastically when different mixing length
algorithms are used.  TYCHO, the stellar evolution code originally
developed by Arnett, is being upgraded to incorporate more realistic
mixing algorithms based on multi-dimensional simulations (Meakin \&
Arnett 2006), but thusfar, the progenitors available are produced by
codes using mixing-length convection.

Other effects might include the initial rotation of the star.  Fryer
\& Heger (2000) found that extreme rotation dampened the convection
along the rotational equator, ultimately leading to a weaker
explosion.  This 15\,M$_\odot$ star ultimately had considerable
fallback, forming a black hole (Hungerford et al. 2007).  Others have
found that rotation can lead to asymmetric neutrino heating that helps
to drive convection and ultimately a supernova explosion (Shimizu,
Yamada \& Sato 1994; Kotake, Yamada, \& Sato 2003).  Also, as we shall
discuss below, Yoon \& Langer (2005) found that rapid rotation could
lead to extensive mixing that burns most of the hydrogen envelope into
helium, producing larger and denser cores that are more likely to
collapse directly to black holes.  Both of these effects will lower
the limiting mass for black hole formation (both through direct
collapse and through fallback).

Let's summarize what we have learned.  Because nearly all of the
progenitors currently suggested require the formation of a black hole,
all predict that it should be easier to form GRBs at lower
metallicities where weaker winds allow more massive
cores.  If a progenitor requires the direct collapse of the star's
core into a black hole (most of the currently proposed progenitors do
not distinguish between fallback and direct-collapse black holes),
current models suggest that GRBs will not occur at solar metallicity.
These models also predict that, if we are limited to direct-collapse
black holes, only stars above $\sim 25-40$\,M$_\odot$ (depending on
the choice of stellar evolution code) will produce GRBs.  Lastly, the
biggest uncertainty in such calculations lies in our poor
understanding of stellar evolution, and it is unlikely that we will
constrain GRB progenitors beyond this current state until real
progress is made with these models.

{\em Angular Momentum:}  

With respect to angular momentum, the progenitors for GRBs can be
divided into two classes: those that are born rotating rapidly and
retain enough of their birth angular momenta to produce a black hole
accretion disk, and those that are spun up through interaction with
another star (either tidal forces or merger events).  Achieving
sufficiently high angular momenta in the collapsing cores is
potentially the strongest constraint on any progenitor.
Unfortunately, stellar evolution models studying angular momentum are
still primitive: they generally neglect centrifugal forces (which can
be important for the high spin rates required to make GRBs), they
incorporate recipes for the generation and angular momentum transport
effects of magnetic fields that may or may not be accurate, and they
depend sensitively on the loss of angular momentum through winds
(Hirschi et al. 2005; Woosley \& Heger 2006; Meynet \& Maeder 2007).
The current state of affairs is that stars with very fast initial spin
periods may retain enough angular momentum to produce an accretion
disk if the mass loss is sufficiently low.

For many binary progenitor models, the binary component is used to
remove the hydrogen envelope without the angular momentum loss that
occurs in wind mass-loss.  But some binary models have been proposed
that use the binary to inject angular momentum into the star at later
stages in the star's evolution.  The helium-merger model of Fryer \&
Woosley (1998) argues that a compact star (either neutron star or
black hole) merges with its companion, injecting angular momentum as
it spirals into the companion's core.  This model definitely will add
angular momentum, maybe even too much (Di Matteo et al. 2002; Fryer et
al. 2006a).  

{\em Ejecting the Hydrogen Envelope:}

The ejection of the hydrogen envelope can occur either through stellar
winds or binary mass ejection.  For single stars, where we must rely
upon stellar winds, this allows us to place constraints on the
metallicity of progenitors.  Heger et al. (2003) found that only a
fraction of single stars will both remove their hydrogen envelopes and
collapse to form black holes (they did not consider constraints to fit
the rotation requirements).  Figure 6 shows GRB rate as a function of
metallicity using the results of Heger et al. (2003) assuming all
stars have sufficient angular momentum to form a disk.  Note that the
peak GRB rate occurs near 0.4\,Z$_\odot$.  The decrease above this 
metallicity value occurs because fewer and fewer stars collapse to 
form black holes.  But the decrease below the peak metallicity value 
occurs because fewer and fewer stars lose their hydrogen envelopes.

As we shall see below, binary models avoid this constraint by
definition: binary models all invoke a mass transfer phase that
removes most or all of the hydrogen envelope.  The Yoon \& Langer
single-star models also avoid this constraint.  But there is a
growing belief that GRBs lose their helium envelopes as well (the
supernovae associated with GRBs do not exhibit strong helium lines).
If so, the Yoon \& Langer single-star model is ruled out.  Without a
better understanding of winds, we can not say much more about this
constraint other than the fact that it will be more restrictive for
single-star models.

\subsubsection{Single-Star Models}

Single-star models can be grouped into two types of progenitors: a
single star with strong winds that eject the entire hydrogen envelope
(Fryer, Woosley, Hartmann 1999), and a single star with extensive
mixing that burns most of the hydrogen to helium (Yoon \& Langer 2005;
Woosley \& Heger 2006).  The high mass-loss case argues that the
progenitors arise from a subset of Wolf-Rayet stars with enough mass
at collapse to form black holes and enough angular momentum to form a
disk.  Unfortunately, most massive stars that lose their hydrogen
envelopes also lose so much mass that they don't collapse to form
black holes.  Using the models of Heger et al. (2003), we find that
the stars that both lose their hydrogen envelopes and still collapse
to form a black hole lie within a narrow range of masses depending on
metallicity: $\sim 32-40$\,M$_\odot$ at twice solar metallicity, $\sim
34-60$\,M$_\odot$ at solar metallicity, $\gtrsim 36$\,M$_\odot$ at
roughly 1/10th solar metallicity, and $\gtrsim 60$\,M$_\odot$ at
roughly 1/1000th solar metallicity.  The fraction of stars forming GRB
progenitors peaks at metallicities around 1/10th solar for this
progenitor scenario.  However, single-star stellar evolution models
including rotation have had trouble getting enough angular momentum in
the core to produce GRBs (Woosley \& Heger 2006).  Winds can
significantly reduce the angular momentum (Hirschi et al. 2005),
making achieving the required angular momenta in the core very
difficult for this progenitor.  Indeed, we expect faster rotating
cores in those stars that retain their hydrogen envelopes, predicting
a larger population of hydrogen-rich hypernovae than the observed
hydrogen-poor class.  We term this progenitor the ``classic single
star'' scenario.  The primary uncertainty in this calculation lies in
the angular momentum transport present in the stellar evolution
models.

An alternate single-star model has recently been proposed by Yoon \&
Langer (2005) and Woosley \& Heger (2006) where a rapidly rotating
core can develop extensive mixing, burning nearly the entire hydrogen
envelope into helium.  Such a star becomes a helium star not by
ejecting its hydrogen envelope, but by burning the hydrogen into
helium.  Such a model will only work with fast-rotating cores at low
metallicities (less than 1/10th solar).  But this constraint is
actually a constraint on the mass-loss rate, and it may be that the
maximum metallicity that can be accomodated is higher.  High core
rotation rates can be attained for these stars and the large helium
core masses lead to larger cores, which are more likely to collapse to
black holes.  This model predicts no GRBs to be produced at
metallicities above 1/10th solar.  The low metallicities also argue
for weak winds.  We do not expect any hydrogen-rich hypernovae or many
helium-poor hypernovae from this progenitor.  We term this the
``mixing single-star'' scenario.

\subsubsection{Binary Mass Transfer Models:}

One way to avoid the problem of loss of angular momentum in a stellar
wind is to eject the hydrogen envelope via binary mass transfer
(Fryer, Woosley, Hartmann 1999).  An example of such a progenitor is a
binary system where, when the more massive star evolves off the main
sequence, it expands and envelops its companion.  The companion then
spirals in toward the core of the massive star, ultimately ejecting
the envelope of the massive star (turning it into a helium star) and
producing a close binary system.  In this simple case, the effect of
the binary is only to eject the hydrogen envelope.  The fraction of
these systems that form hypernovae depends on the fraction of stars
that are in close binaries (presumably lower at low metallicity
because low metallicity stars don't expand as much\footnote{This trend
  is not completely accepted in the stellar community.}), the fraction
of stars that collapse to form black holes (larger at low
metallicity), and the angular momentum evolution of massive stars.  It
is likely that the fraction goes up with decreasing metallicity, but
stellar evolution models are not at the level that they can answer
these questions yet.  Some bursts will have strong winds, but it is
likely that more will have weak winds (strong winds will lead to the
formation of neutron stars, not black holes).  None of these
progenitors will be hydrogen-rich and some will be helium poor.  We
term this model the ``classic binary'' scenario.

A subset of these binaries will produce such tight binaries after the
mass transfer phase that the two stars will become tidally locked (as
suggested by Izzard et al. 2004; see also Tutukov \& Cherepaschuck
2004). In this manner the binary would not only remove the hydrogen
envelope but also spin up the massive star. If the angular momentum is
conserved through the collapse, it may be more than sufficient to form
a black hole accretion disk.  To determine whether such a mechanism
can work, two issues must be tested.

First, we must ensure that tidal synchronization is sufficiently fast
for it to occur in our quickly-evolving stars.  Yoon \& van den Heuvel
(2006,2007) calculated these timescales for helium stars. They found
that if the post-inspiral companion of the helium star is a
Roche-lobe-filling solar-type main sequence star (typical orbital
periods around 6 to 8 hours) the helium star reaches tidal
synchronization in a fraction of its lifetime because of magnetic
coupling between core and envelope (using Spruit's 2002 mechanism).
It is likely to remain close to solid body rotation until the final
core collapse. In that case it has at the time of core collapse too
little angular momentum to produce a GRB.

Second, we must test whether tides spin up the star fast enough to
produce a GRB.  Yoon \& van den Heuvel found that if the companion of
the helium star is itself a compact object, the shortest possible
orbital periods are $\sim$1 hour and in this case, with the same
prescription the core has sufficient angular momentum to make a GRB.
Their results therefore suggest that only compact binary systems that
descended from High Mass X-ray Binaries (HMXBs) would be able to
produce GRBs. They argue that several well-known galactic HMXBs, such
as Cyg X-1 and 4U1223-62 are excellent prospective progenitors of
these very close helium-star-plus-compact-star binaries, and that
systems of this type may be produced in sufficiently large numbers to
make a sizeable contribution to the long-duration GRB formation rate.
The preference of long-duration GRBs for small, lower-metallicity,
star-forming galaxies (see below) would then be due to the lower wind
mass-loss rates in low-metallicity massive stars (e.g. see Lamers \&
Casinelli, 1999, Mokiem 2006), which favors black-hole formation
over neutron-star formation at stellar collapse.

Belczynski et al. (2007) discovered this same result by studying a
series of progenitors with and without this tidal locking effect.  As
a base model, they assumed no tidal locking and a standard progenitor
with reasonably fast rotation at birth, but included the mechanism by
Spruit (2002) for magnetic braking.  For their tidal calculations,
they used this same base model, but for those stars where the radius
of the star was greater than $0.2a(1-e)$ where $a$ is the orbital
separation and $e$ is the eccentricity, they assumed that the star was
completely synchronized.  Although a population of the secondary stars
are spun up (the previously-mentioned HMXBs), the bulk of the close
binaries are slowed down by tidal effects.  Those that are spun up are
quite rare and unlikely to match the observed GRB rate.

\subsubsection{Binary Merger Models:}

If we include magnetic braking (using Spruit's 2002 mechanism) in
stellar evolution models, we find that neither of the binary
mass-transfer models currently have core angular momenta at collapse
that are sufficient to produce collapsars (Woosley \& Heger 2006).
This leads one to suggest increasingly exotic progenitors.  One such
progenitor argues that the two stars in the binary have nearly equal
masses and hence the companion evolves off the main sequence before
the more massive star collapses (so that the binary goes through two
common envelope phases prior to collapse).  In the second common
envelope phase, the stars merge, producing a single massive star which
has lost most of the hydrogen envelopes of both stars (Fryer et
al. 1999).  The merger process injects much of the orbital angular
momentum of the binary into the merged star, providing considerable
spin-up with a nearly bare helium star.  It is assumed that winds will
remove the rest of the envelope.  Simulations have shown that if
neither star has begun helium burning before the merger, the final
collapsed core will be spinning slightly faster than its single star
counterpart (Fryer \& Heger 2005).  It is likely that if the more
massive star is well through helium burning, this spin-up will be more
dramatic, but simulations to confirm this trend have yet to be done.
This progenitor will have similar observational trends to the binary
mass transfer scenarios.  Initially named the Helium-Helium merger
scenario by Fryer et al. (1998), this name has produced incredible
confusion with the He-merger model below, so in this paper, we will
rename it.  In deference to the Bethe \& Brown (1998) proposal to use
equal-mass stellar binaries to make double neutron star binaries, we
term this the ``Brown Merger'' scenario.

An additional progenitor scenario based on the new route for
common-envelope ejection was discovered by Ivanova and collaborators
(Ivanova \& Podsiadlowski 2003; Podsiadlowski et al. 2006).  They
noticed that in Case C common envelope phases, the inspiraling
secondary star can actually overfill its Roche-lobe and accrete onto
the helium core of the primary.  This accreting material will affect
the core in two ways.  First, this mass accretion will spin up the
core.  Second, the material streaming from the secondary can penetrate
deep into the helium core and ignite, producing explosions that eject
not only the helium shell but the hydrogen envelope as well (Ivanova
\& Podsiadlowski 2003).  The final product of this explosive mass
ejection is a pure CO core, consistent with current observations of
the supernovae associated with GRBs.  Since this scenario only occurs
in Case C mass transfer, the GRB will occur shortly after (within
$10^4$\,y) the explosive mass ejection and the shell from this common
envelope ejection should still be relatively close (within roughly a
parsec).  We will term this scenario the ``Explosive Ejection''
scenario.

\subsubsection{He-merger Model}

Fryer \& Woosley (1998) proposed a model akin to the bulk of the
collapsar models arguing that the merger of a neutron star or black
hole with its companion could produce a collapsar-like outburst.  In
this formation scenario, the binary first evolves into a neutron star
or black hole binary (observed as X-ray binaries).  In many
conditions, the companion eventually envelopes the the compact object,
causing it to spiral into the center of the companion star.  This
progenitor avoids the difficulties involved in forming black holes,
and it easily spins up the collapsing star enough to form a disk.
However, it may have too much angular momentum (Di Matteo et al. 2002;
Fryer et al. 2006a).  Compared to the mass transfer and single star
scenarios, it is not so strongly dependent on the metallicity.  But
since the binary is likely to be kicked in the formation of the
compact remnant, the binary can move significantly beyond its
formation site and may not be enshrouded in a stellar wind (it will,
however, have a torus of ejected envelope material in the equator of
the rotation axis and GRB jet).  Very few of these progenitors will be
hydrogen-rich, but one would expect the bulk of them to be
helium-rich.  This is the ``He-merger'' scenario.

One way to overcome the angular momentum problem is to assume that the
common envelope phase occurs after helium burning (Case C mass
transfer).  In this case, the moment of inertia of the C/O core will
be larger, and the inspiralling neutron star will have lost much of
its orbital angular momentum, leading to slower rotation.  If only
these mergers lead to a collapsar engine, a large fraction may be
helium-poor.  Otherwise the observational properties of this subclass
of He-mergers is similar to the classical He-merger scenario.  We term
this the ``Helium Case C'' scenario.

\subsubsection{Cluster Models}

In this workshop, it was proposed by Kulkarni that perhaps the
progenitor requires interactions in a cluster.  Two possibilities for
such progenitor include cluster enhanced mergers, currently invoked to
form intermediate mass black holes (Portegies-Zwart et al. 2005) or
mergers between compact remnants and stars.  These mergers may well
produce massive, rapidly spinning cores.  They also will be more
common in low-metallicity systems.  Thusfar, no detailed studies have
been done on these systems.

\section{Observational Constraints}

By 1992, the number of gamma-ray burst models proposed by theorists
had grown to over 100 (Nemiroff 1994).  Although many of these models
stretched the limits of physics, the bulk were only discarded when it
was shown that the models did not match observations (or did not match
the observations as well as other models).  Rightly so, observations
are used as the final arbitrator in theoretical disputes and have
played a major role in our understanding of GRB and SN progenitors.
Here we discuss those observations that can be used to constrain GRB
and type Ib/c progenitors.  For each constraint, we outline both the strong 
result (what we believe is robust) and the trend (what is implied by 
the data).  These results are summarized in table~\ref{table:observation}.

\subsection{Rates}

As theorists introduce increasingly exotic progenitors for GRBs, the
rate of these bursts becomes an ideal constraint on the models.  Any
proposed model must be able to produce bursts at a rate comparable to
the observed rate.  Generally, the mode of operation is to make sure
that, under optimistic conditions, the GRB rate is larger than the
observed rate.  The reference value for the observed rate of GRBs per
average galaxy has been estimated from the BATSE monitoring as
$R_{\rm obs} \sim 10^{-7}$ yr$^{-1}$ (e.g. Zhang \& Meszaros 2004) of
which 2/3 are long-duration GRBs. However, GRBs are highly beamed and
can be detected only if the observer is within the small jet opening
angle. This implies that the intrinsic GRB rate is likely a factor
$10-100$ higher, $R_{\rm true} \sim 10^{-6}-10^{-5}$ yr$^{-1}$
galaxy$^{-1}$ (Podsiadlowski et al. 2004). There is evidence that the
rate rapidly increases with redshift and is $10-100$ times higher
already at redshift $z=1$ (Firmani et al. 2004, Matsubayashi et al.
2005).

The number above applies to the average or ``normal'' GRBs seen at
cosmological distances. On the other hand, the recent discovery of
underluminous, relatively nearby XRFs and GRBs suggests the existence of
a population of events less luminous but possibly $10^2$ times more
frequent (Pian et al. 2006). If these events are considered, the local
rate of GRBs may be as high as $\sim 10^{-4}$ yr$^{-1}$ galaxy$^{-1}$.

The rate of all core-collapse SNe in the local Universe is $6 \times
10^{-3}$ yr$^{-1}$ galaxy$^{-1}$ (H$_0$ = 70 km s$^{-1}$ Mpc$^{-1}$).
According to the latest published estimates, still based on
photographic/visual SN searches, type Ib/c account for 15\% of all
core-collapse SNe, i.e. $\sim 10^{-3}$ yr$^{-1}$ (Cappellaro et al.
1999) although preliminary analysis of modern CCD SN searches suggests
that this number may need to be increased by a factor $\sim2$.  About
5--10\% of the observed SNIb/c, showing high expansion velocity and
bright luminosity, are usually dubbed hypernovae (Podsiadlowski et al.
2004, Richardson et al. 2006).

It has been found that the rate of core collapse also rapidly
increases with redshift (Dahlen et al. 2004, Cappellaro et al. 2005)
closely tracking the star formation history. This is consistent with
the notion that the progenitors of both GRBs and core collapse SNe are
massive stars. At the moment there is no information on a possible
evolution with redshift of the specific SNIb/c or hypernova rate.  

Even allowing for the large uncertainties of current GRB statistics,
it is safe to conclude that only a small fraction ($\sim 1-10\% $) of
SN Ib/c can be associated with GRBs, a fraction which appears to
coincide with that of hypernovae.  This corresponds to roughly
$\sim$1\% of all core-collapse supernovae being GRBs.  Such estimates 
are corroborated by radio surveys of supernova remnants (Soderberg 
et al. 2004; Gal-Yam et al. 2006).

Theory estimates that roughly 5-40\% of all core-collapse stars
collapse to form black holes (Fryer \& Kalogera 2001).  The primary
uncertainty in this fraction comes from uncertainties in the initial
mass function of massive stars.  If $\sim$1\% of all core-collapse
stars produce GRBs, this means that it may be that 20\% of all
black-hole forming stars must form GRBs.  If the number were indeed
this high, many of the current progenitors would be ruled out.
However, with the current uncertainties in the rates and the initial
mass function, this value could be as low as 1\%, well within the
range of the progenitors proposed here.  But as the data and the
theory behind the progenitors become more firm, it is likely that
rate estimates will be able to rule out certain models.

The relative rates of type Ib/c and type II supernovae and GRBs may
also rule out some of the progenitors.  Some of the binary and single
star models make very different predictions for the metallicity
dependence, and to a lesser extent redshift dependence, of the type
Ib/c to type II supernova ratio (compare the results shown in Figs. 1
and 2).  Reliable ratio values, discussed further below, could well
rule out many of the models.

\subsection{Supernovae Associated with Gamma-Ray Bursts}

Collapsing massive stars that lose their hydrogen envelope are compact
stars.  Although their explosion mechanism is very different from type
Ia supernovae, they have the same rapid light-curve evolution and
absence of hydrogen lines seen in type Ia supernovae and therefore are
classified as type I supernovae.  Initially, type I supernovae were in
one class, but as more type I supernovae were discovered with spectral
appearances very different from standard I supernovae, the new
classifications of type Ib, and later Ic, were introduced (Filippenko
1997).  In particular, SNe Ib have strong He lines in their spectra.
These He lines strongly suggest that the origin of such supernovae are
massive stars that have lost their hydrogen envelopes.  Helium lines
are notoriously difficult to excite. In a classical paper, Lucy (1991)
showed that the high HeI levels responsible for the optical lines
cannot be significantly populated by thermal mechanisms at the
temperatures typical of SN atmospheres.  The most efficient mechanism
is non-thermal excitation/ionization by the fast particles produced by
the diffusion of the $\gamma$-rays and the positrons emitted in the
decay of $^{56}$Ni into $^{56}$Co and then into $^{56}$Fe. Departure
coefficients of the order of $10^4-10^6$ can be easily attained. The
more $^{56}$Ni that is mixed out into the He layer, the easier it is
for non-thermal processes to take place.  Therefore, seeing strong HeI
lines means both that non-thermal excitation/ionization is strong and
that the mass of He in the ejecta is rather large.  Typically, massive
stars develop He shells with masses above $\approx 1 M_\odot$ (Woosley
et al. 2002, Yoon et al. 2006).  Likely progenitors of SNe Ib are
therefore WR stars, which lose their hydrogen envelope via strong
stellar winds.  Without the new Yoon \& Langer (2005) mixing model or
binary stars, the stars that produce these Ib supernovae must be more
massive than 34\,M$_\odot$ at solar metallicity (Heger et al. 2003).
This limit moves upward at lower metallicities.

A later addition to the SN zoo, SNe Ic are characterized by the
absence of both H and He in their spectra, as well as by the weakness
or absence of the Si and S lines that are more typical of SNe Ia
(again see Filippenko 1997).  The obvious progenitors for SNe Ic are
then stars that have lost both the H and He envelopes. The
pre-explosion star may thus be an early-type WR star, such as a WC
star.  However, it is not clear that such extreme stripping can be
achieved in a single star configuration.  Nomoto et al. (1995)
suggested that SNe Ib come from single stars while SNe Ic come from
binary stars. A binary configuration helps to remove the envelope
(Podsiadlowski et al. 1992), and it may also address the angular
momentum problem (see above).  Actually, observationally there are
many more SNe Ic ($\sim 2/3$) than SNe Ib. This is against intuition
in the single star case, and may be a further argument in favour of a
binary origin for SNe Ic.

Some claims have been made of the presence of He lines in SNe
Ic. These are mostly based on the difficulty in identifying a strong
absorption line that is present near 1$\mu$m in the early-time spectra
of SNe Ic. One possible identification of the line is in fact HeI
10830\AA. A similar situation actually exists for SNe Ia. In this
latter case, Mazzali \& Lucy (1998) showed that there are alternative
possibilities (e.g. SiII, MgI), and that if a strong HeI 10830\AA\
line is present, then a strong 2$\mu$m line is also expected. Infrared
data of SNe Ic (Taubenberger et al. 2006) show that this is not the
case.  Therefore, if there is any He in SNe Ic, it is not a lot.

So far, there are only 4 well-observed cases of SNe associated with
GRBs or XRFs: SN1998bw/GRB980425, SN2003dh/GRB030329,
SN2003lw/GRB031203, and SN2006aj/XRF060218.  All these SNe are of Type
Ic, and they also share a broad-lined spectrum, which is indicative of
the ejection of material at very high velocities $\sim 50000$km/s
(Iwamoto et al. 1998; Mazzali et al. 2003, Deng et al. 2005; Mazzali
et al. 2006; Pian et al. 2006).  Other broad-lined SNe Ic without GRBs
are known (e.g. SN97ef, SN02ap etc.; see Nomoto 2005 for a review).
Broad-lined SNe Ic account for $\sim 5-10$\% of all SNe Ic, and
GRB/SNe account for $\sim 20$\% of all broad-lined SNe Ic. In
contrast, no broad-lined SNe Ib have been observed, let alone in
conjunction with a GRB. These numbers are beyond what the relative
rate of SNe Ib v. Ic would predict, and one may therefore wonder
whether having a SN Ic is a pre-requisite for a) ejecting material at
high velocities, and b) producing a GRB (which is probably the most
extreme part of a) and/or an orientation dependent property)

So any progenitor scenario must produce a reasonable number of
progenitor stars that, at collapse, lose not only their hydrogen
envelopes, but also most of their helium envelopes.  If the current
trend holds, the observations could require that all progenitor 
stars must lose most of their helium envelopes.

\subsection{Metallicity}

A generic characteristic of massive stars is that these objects drive
significant stellar winds.  During the main sequence lifetime of an O
star, the mass loss rate is $\approx 10^{-7}-10^{-6} M_\odot {\rm
yr^{-1}}$ for stars of roughly solar abundance.  The mass loss rate
rises dramatically ($> 10^{-6}-10^{-5} M_\odot {\rm yr^{-1}}$) during
the red giant phase and later stages of the star's life (e.g.\ the
Wolf-Rayet phase).  Because these winds are driven by the radiation
pressure of UV photons on metals in the stellar atmosphere, the wind
mass-loss rates are predicted to be very sensitive to stellar
metallicity (e.g. Nugis \& Lamers 2000; Kudritzki 2000, Vink \& de
Koter 2005).  One of the easier (although still quite difficult)
predictions a progenitor model can make lies in the metallicity
dependence.

For the foreseeable future, we are unlikely to develop an
observational technique which will permit a direct metallicity
measurement of a GRB progenitor.  Perhaps the only prospect is if a
GRB were to occur very nearby and one inferred the stellar metallicity
from the SN ejecta.  In lieu of this approach, observers can infer the
metallicity of gas near the GRB progenitor using a few complementary
approaches: (1) absorption line spectroscopy of GRB afterglows; (2)
emission line spectroscopy of \ion{H}{2} regions within the GRB host
galaxy; (3) the slightly less direct method of measuring the
interstellar extinction in the host (as dust content is related to
metallicity); and (4) the very indirect measurement using the
host-galaxy morphology: if the hosts are small SMC/LMC-like galaxies,
they are likely to have lower metallicity (see below).  The first
approach is currently limited to high-redshift $(z>2)$ GRB events.  At
lower redshifts, measurements of hydrogen via the Ly $\alpha$,
$\lambda 1215$ transition require UV spectra and therefore space-borne
telescopes.  In contrast, the second approach is generally restricted
to low redshift $(z<0.6)$ such that key emission lines remain in the
optical spectrum.  As such, there is no GRB host galaxy where the two
techniques have been compared.

By comparing the column densities of hydrogen versus a metal (e.g.\ O,
S, Fe) one can derive the metal abundance of the interstellar medium
(ISM) surrounding the GRB.  The observations are restricted to
gas-phase abundances and one preferentially focuses on non-refractory
elements (e.g.\ O, S, Zn) to minimize the effects of depletion onto
dust grains.  The spectra of GRB afterglows reveal strong interstellar
absorption lines and damped Ly $\alpha$ profiles (e.g. Barth et
al. 2003; Vreeswijk et al. 2004; Chen et al. 2005).  The \ion{H}{1}
column densities are easily derived from even low-resolution, moderate
signal-to-noise ratio observations by fitting the damping wings of the
Ly $\alpha$ transition (e.g. Jakobsson et al. 2006).  An accurate
determination of metal-line column densities, however, is challenged
by the very large observed equivalent widths (e.g. Savaglio et
al. 2003).  First efforts reported metal column densities based on
traditional single-component, curve-of-growth (COG) analysis
(e.g. Savaglio et al. 2003), yet high-resolution observations of GRB
afterglows indicate that the COG results systematically underestimate
the metal abundance (Prochaska 2006).  In general, one can only
estimate a lower limit to the metallicity for spectra that do not
resolve the line profiles.

Because the majority of observations to date were acquired with
low-resolution spectrometers, there is a preponderance of lower limits
to the metallicity (Prochaska 2006).  In any case, the present set of
metallicity measurements from $\approx 10$ GRB afterglow spectra
exhibit a large dispersion of values from $\approx 1/100$ solar
(Chen et al. 2005) to nearly solar metallicity (Castro et al. 2003).  Many of
the lower limits lie at $\approx 1/10$ solar metallicity (Prochaska 2006)
and, therefore, the mean (or median) value is at least this enriched.
Furthermore, an average metallicity of 1/3 to 1/2 solar is permitted
if not suggested by the data.  Even adopting the lower limits as
values, the distribution currently lies along the upper threshold of
damped Ly $\alpha$ metallicity measurements along quasar sightlines
(Prochaska 2003).  That is, the ISM measurements for GRB host galaxies
match, and likely exceed, the cosmological mean metallicity in neutral
gas at $z>2$.  In this respect, at least, the GRB have average or even
super-solar values.  The full distribution of GRB metallicities from
afterglow spectroscopy awaits the compilation of a much larger sample
of echelle observations.

Apart from studying absorption and emission lines, one also may obtain
information on the metallicity by measuring the interstellar
extinction by dust in the host galaxy. Here all investigations find
remarkably low dust contents in GRB hosts, the simplest interpretation
of which is low metallicity. The interstellar extinction curves are
very different from that of our galaxy. None of the hosts shows the
2175\AA\, extinction bump. The extinction curves of the hosts resemble
more that of the SMC, which has a much higher gas to dust ratio than
our galaxy. An important recent investigation by Starling et al.
(2006) finds that, in all cases where the gas to dust ratio of hosts
can be determined, it is equal to or larger than that of the SMC.

Before concluding our discussion of ISM metallicities, we wish to
comment on two easily overlooked aspects of the measurements: (i) the
relation of the observed gas to the GRB progenitor; and (ii) the
abundance of Fe.  Although one may expect the GRB progenitor to reside
within a molecular cloud and/or to be surrounded by circumstellar
material, afterglow spectra have not revealed strong evidence for this
gas (Prochaska et al. 2006; Chen et al.2007).  Furthermore, the GRB
afterglow spectra almost always show strong \ion{Mg}{1} absorption
which must occur at a distance $>50$\,pc from the afterglow to avoid
photoionization (Prochaska et al. 2006).  Similarly, Vreeswijk et
al. (2006) infer a distance to the ISM of $\approx 1$\,kpc based on
their analysis of varying Fe$^+$ and Ni$^+$ fine-structure levels.
These observations, therefore, indicate that the majority of neutral
gas along the GRB sightline is at 100pc to 1kpc distance.  In turn,
the metallicity measurements must be considered at best a crude
estimate for the GRB progenitor.  The second point to emphasize is
that the absorption-line measurements do not give a precise
measurement of Fe or any other element on the Fe-peak.  This is due to
their refractory nature; these elements are easily depleted from the
gas-phase onto dust grains.  One can set a lower limit to the Fe-peak
abundances from gas-phase measurements, but the corrections for
differential depletion can exceed an order of magnitude.  The
metallicity values described above correspond solely to S, Si, and Zn,
none of which dominate the opacity in massive stellar atmospheres.  As
it is reasonable to assume that the gas has an abundance pattern
typical of massive star nucleosynthesis (i.e.\ $\alpha$-enriched), the
Fe abundance is likely $\approx 2$ times lower than that recorded for
Si or S.

Turning to lower redshift, one can infer the metallicity of the GRB
progenitor by analyzing forbidden emission lines from \ion{H}{2}
regions within the host galaxy.  Aside from very local galaxies, the
observations generally contain emission from the entire galaxy.
Nevertheless, one does not tend to observe very large metallicity
gradients in \ion{H}{2} regions (Kobulnicky 2005) and the derived
value should correspond (within a factor of 2) to the local \ion{H}{2}
region.  The analysis involves standard techniques of comparing line
fluxes of forbidden H, O, and N transitions against models of
\ion{H}{2} regions.  To date, the results (which have been limited to
GRB host galaxies at $z < 0.5$) reveal values of $\approx 1/10$ solar
metallicity (Prochaska et al. 2004; Sollerman et al. 2005).  Stanek et
al.(2006) have used these measurements to argue that low $z$ GRB host
galaxies are biased to low metallicity under the assumption that GRB
trace current star formation.  That is, a random sample of star
forming galaxies (weighted by current SFR) in the local universe would
have higher average metallicity than that observed for the GRB host
galaxies.  At present, the small sample size precludes a strong
conclusion regarding metallicity but the results are suggestive of a
selection bias.  We also caution that the $z<0.5$ GRB events have
systematically lower energy than $z > 1$ events and one should not
generalize these results to cosmological bursts.

Not surprisingly, the low $z$ GRB host galaxies also have low
luminosity.  Furthermore, Kewley et al.(2006) note that the GRB host
galaxies also fall off the luminosity-metallicity (L-Z) trend observed
for other low $z$, irregular galaxies.  The offset is in the sense
that GRB host galaxies are either especially luminous for their
metallicity or metal-poor given their luminosity.  Given that GRBs are
associated with short-lived massive stars, a possible explanation for
the offset is that the star formation leads metal enrichment.  That
is, the galaxy is exhibiting a burst of star formation yet has not had
sufficient time to enrich its \ion{H}{2} regions.  This is consistent
with the observation that GRB host galaxies at $z\sim 1$ have very
high specific star formation rates (SFR normalized to the galaxy
luminosity; Christensen et al. 2004).

Finally, indirect evidence concerning the metallicities of long GRB
environments may come from the study of the morphology of their host
galaxies. Already several years ago it was noticed that the long GRB
hosts tend to be small galaxies that are systematically bluer than the
same size galaxies in the general population at similar redshifts.
This suggests that they contain many massive stars and have a larger
SFR than the general population of galaxies of similar size. Still,
only a few GRB hosts show evidence of strong starbursts. It is not
clear therefore that GRBs can be used to directly trace current SFR.
A systematic, comparative study by Fruchter et al. (2006) compares 42
GRB host galaxies observed with HST against the host galaxies of
serendipitously HST-discovered core-collapse supernovae (CC-SN; types
Ib/c and II) at the same general redshift ($z < 1.2$).  This
comparison shows striking differences between the two host galaxy
populations and between the localizations of the explosions relative
to the galaxies' light distributions.  Specifically, of the 42 long
GRB hosts, 41 appear to be small star-forming galaxies; only one of
them is a Grand Design Spiral, whereas half of the hosts of the CC-SN
at similar redshifts are Grand Design Spirals.  In addition, the
locations of the GRBs on their host galaxies appear to be strongly
concentrated on the optically brightest parts of their hosts, whereas
the CC-SN follow the average light distribution of their hosts.  The
bright spots on the light distribution of the GRB hosts are typically
large concentrations of massive young stars (starbursts), similar to
those observed in nearby blue dwarf galaxies such as NGC 3125 , a
galaxy with metallicity similar to that of the LMC/SMC (Hadfield \&
Crowther 2006), with a blue clump containing on the order of ten
thousand O and Wolf-Rayet stars.  Since the sizes of the GRB hosts are
typically like those of the LMC and SMC, which have metallicities 0.3
and 0.2 solar, respectively, the morphology of the GRB hosts and the
localizations of the GRBs on their hosts are therefore consistent with
the idea that long GRBs occur at lower-metallicities than their normal
supernova counterparts.  Results like those of Modjaz et al. (2007)
corroborate these results.  All of the progenitors listed in this
paper require lower metallicities than normal supernovae.  But what we
mean by lower metallicity is still a subject of debate.  Wolf \&
Podsiadlowski (2007) use the Fruchter et al.(2006) GRB host galaxy
sample and show statistically that the median GRB host galaxy is a
galaxy with the mass of the LMC and a metallicity of half solar.  
Indeed, they even argue that any model that requires a metallicity 
well below half solar can effectively be ruled out.

As to the normal CC-SNe: due to the shape of the IMF one expects some
75\% of all CC-SNe to originate from stars in the mass range 8--20
M$_\odot$, where core-collapse produces neutron stars. The bulk of the
CC-SNe are therefore expected to be neutron-star-forming events.  The
striking difference between the morphologies of the long GRB hosts and
the CC-SN hosts, as well as their differences in localization with
respect to hosts light distribution, therefore strongly suggest that
long GRBs are different from neutron-star forming supernovae. As
pointed out by Fruchter et al. (2006) this is consistent with the
picture that we are dealing here only with collapses of the cores of
the most massive stars, which collapse to a black hole (or possibly
other phenomena related only to the most massive stars).

At this point in time, it appears that there is no consistent picture
of the metallicities of GRBs.  We note however that direct
measurements argue for higher metallicities, whereas the indirect
measurements suggest lower metallicities.

\subsection{Surrounding Environment}

Some relevant properties of the medium around GRB progenitors can be extracted 
(at least, in principle) from observations of GRB afterglows in two ways.
The more direct way is provided by understanding the origin of the absorption features 
seen in the afterglow optical spectrum. The less direct way is to compare the observed 
afterglow light-curve with the analytical expectations for the blast-wave model, with 
the aim of constraining only the most generic properties of the burst ambient medium.

{\em High-Velocity Absorption Lines in Afterglow Spectrum}

High velocity absorption lines of CIV and SiIV, blueshifted by 450,
1000, and 3100 km/s have been identified in the optical spectrum of
the GRB afterglow 021004 (Mirabal et al 2002, Schaefer et al
2003). CIV and lower ionization species (FeII, AlII, MgII) absorption
lines have also been seen in the spectrum of GRB afterglow 020813
(Barth et al 2003), at an outflowing velocity of 4300 km/s, and in
that of GRB afterglow 030226, blueshifted by 2300 km/s. These
velocities are too high to arise in a galaxy cluster, i.e. the
absorbers must be in the GRB host galaxy. It cannot be ruled out
definitely that the absorbers are the outflow of a now-dormant QSO or
a superwind from a starburst region but these origins are unlikely,
given the required GRB direction -- QSO outflow chance alignment and
the lower velocity measured for starburst winds.  If we can
show that the lines must arise from the progenitor itself, we
have a potentially strong constraint on the mass-loss and collapse
mass of the progenitor (e.g. van Marle et al. 2005).

{\em Circumburst Medium Constraints from Afterglow Light-Curves}

The afterglow emission is believed to arise from the medium within 1
pc of the burst, which is energized by the relativistic shock driven
by the GRB ejecta. The mechanism which generates magnetic fields of
order 1 G in the shocked gas and accelerates electrons to at least 100
GeV (in the comoving frame) is not well-understood but such conditions
must be met by the blast-wave to radiate synchrotron emission at X-ray
energies for days after the burst.

The decay of the afterglow light-curve is determined by the dynamics
of the forward shock which energizes the burst ambient medium. Shortly
after the burst, the blast-wave becomes quasi-adiabatic
(i.e. radiative losses are negligible) and, if there is no energy
injection into the forward shock, the shock dynamics is determined
only by the radial structure of the ambient medium and the collimation
of the GRB outflow.  Before collimation starts to affect the afterglow
dynamics, the radius of the blast-wave is $R_a (t) = 0.25\,
(E_{52}/n_0)^{1/4} [t_d/(z+1)]^{1/4}$ pc, where $E_{52}$ is the
shock's kinetic energy per solid angle in units of $10^{52}$ erg/sr
(i.e. of the order of the GRB output), $n_0$ is the proton density at
the location of the blast-wave, and $t_d$ is the observer time
measured in days.

Therefore the forward shock is within 1 pc of the GRB progenitor for
the entire duration of the afterglow observations. The free winds of
WR stars extend over 10 pc (Castor, McCray \& Weaver 1975,
Garcia-Segura, Langer \& Mac Low 1996) thus the GRB ejecta should
interact with the WR free wind. Then one expects that the afterglow
light-curve ``reflects'' its $r^{-2}$ radial stratification. This
simple test appears straightforward given that the afterglow model
(e.g. M\'esz\'aros \& Rees 1997) predicts a simple linear relationship
between the exponent $\alpha$ of the afterglow light-curve power-law
decay ($F(t) \propto t^{-\alpha}$) and the exponent $\beta$ of the
afterglow power-law spectrum ($F_\nu \propto \nu^{-\beta}$): $\alpha =
1.5\,\beta + c$, the stratification of the circumburst medium, ${\rm
d}\log n/{\rm d}\log r$, setting the coefficient $c$.

There are two complications with the above test for the circumburst
medium stratification.  First, the coefficients of the relationship
between $\alpha$ and $\beta$ change if there is energy injection into
the blast-wave energy and if the microphysical parameters that
quantify the post-shock energy in magnetic field and relativistic
electrons evolve, which can hide the signature of the circumburst
medium in the $\alpha-\beta$ relation.  The X-ray light-curves of
Swift afterglows indicate that there may be a sustained energy
injection in the blast-wave for hours after the burst (Nousek et al
2006, Panaitescu et al 2006, Zhang et al 2006).  The second
complication is that, at observing frequencies above the cooling
frequency ($\nu_c$), the afterglow emission arises from the medium
which was swept-up by the blast-wave within less that one dynamical
timescale, leading to a light-curve decay index $\alpha_x$ which is
independent of the medium stratification. It is quite likely that the
X-ray domain lies above $\nu_c$, thus the decay of X-ray light-curves
cannot constrain the circumburst medium stratification, as illustrated
in fig. 8 of De Pasquale et al (2006). The optical domain is more
likely to be below $\nu_c$, but a good determination of the optical
spectral index $\beta_o$ requires accurate near-IR measurements, to
have a sufficiently wide frequency baseline and to correct $\beta_o$
for dust reddening in the host galaxy.

For a sample of two dozens pre-Swift GRB afterglows with optical decay
indices and spectral slopes measured at about 1 day, the above $\alpha
- \beta$ test identifies 3 cases which require a wind-like medium and
5 for which the medium should be homogeneous; for the rest, the
uncertainties of $\alpha$ and $\beta$ are sufficiently large that
either type of medium is allowed.

Another estimator of the ambient medium stratification results from
comparing the optical and X-ray decay indices, $\alpha_o$ and
$\alpha_x$. If the cooling frequency $\nu_c$ is not between optical
and X-ray then $\alpha_o = \alpha_x$ and no information can be
obtained in this way about the medium stratification. If $\nu_c$ is in
between optical and X-ray then $\alpha_o < \alpha_x$ for a homogeneous
medium and $\alpha_o > \alpha_x$ for a wind. The different decay
indices are caused by the evolution of $\nu_c$: it decreases for a
homogeneous medium and increases for a wind. This test is more robust
because it relies only on light-curve decay indices, which can be
measured more accurately than spectral slopes, and because energy
injection in the blast-wave speeds up the evolution of $\nu_c$, which
increases the difference between the optical and X-ray decay indices,
i.e. the test is not ``spoiled'' a potential departures from the
standard forward-shock model.

There are nearly three dozens of GRB afterglows with measured decay
indices $\alpha_o$ and $\alpha_x$ within the first day after trigger,
only 4 of them exhibit the $\alpha_o > \alpha_x$ expected for a wind
medium while for 10 afterglows $\alpha_o < \alpha_x$ indicate a
homogeneous medium; the rest provide an inconclusive test for the
ambient medium.
 
A third method of assessing the structure of the burst ambient medium
using afterglow observations applies to those afterglows whose
power-law decaying light-curves exhibit a steepening. Such a
steepening is observed in the optical emission of pre-Swift afterglows
at about 1 day and is most likely due to the collimation of the GRB
ejecta, a light-curve break resulting when the blast-wave has
decelerated enough that the emission from the jet boundary becomes
visible. A tight collimation of the GRB outflow, into a jet of
half-opening less than $10^{\rm o}$, is also desirable on energetic
grounds, as the isotropic-equivalent of many GRBs exceeds $10^{53}$
ergs, the output of some burst being even higher than $10^{53}$
ergs. However, we note that the X-ray emission of Swift afterglows
(which have been followed less systematically in the optical) rarely
exhibits a $\sim 1$ day break consistent with a jet origin (Willingale
et al 2007, Sato et al 2007).

The existence of a jet-break in the afterglow light-curve allows us to
distinguish between a homogeneous and a wind-like medium through that
the slightly faster deceleration produced by the former leads to a
shorter time during which the jet edge becomes visible and to a lower
lateral spreading of the jet during this transition phase. The effect
is a sharper light-curve break for a homogeneous medium than for a
wind (Kumar \& Panaitescu 2000). This is true also for a light-curve
break resulting when the symmetry axis (of maximal ejecta kinetic
energy per solid angle) of an angularly non-uniform outflow becomes
visible to the observer (Panaitescu \& Kumar 2003). Numerical
calculations of the jet dynamics and its synchrotron emission are
required to compare the shape of the light-curve break produced by a
jet (or a structured outflow) and observations.

Table \ref{jets} lists the reduced $\chi^2$ obtained for 9 pre-Swift
afterglows whose optical light-curves exhibited a break. These
afterglows have been monitored also at radio and X-ray frequencies, a
multiwavelength afterglow coverage being necessary to constrain the
blast-wave dynamical parameters which determine the jet Lorentz factor
and lateral spreading during the light-curve steepening. Most of the
best fits to the afterglows of Table \ref{jets} are not statistically
acceptable; often, the large $\chi^2$ is due to small scale variations
in the afterglow light-curve, which the model cannot reproduce. In
general, systematic differences between model and observed
light-curves are seen only for $\chi^2 > 4$.  As can be seen from
Table \ref{jets}, only one afterglow is fit better with a wind-like
medium while six are accommodated better by a homogeneous medium. The
score is basically the same for the structured outflow model.

The conclusion that can be drawn from the above is that both the
analytical and the numerical analysis of afterglow light-curves
indicate that the medium into which the GRB ejecta runs has a uniform
density within the first parsec. A wind-like medium, as expected
around WR stars up to 10 pc, is sometimes compatible with the
afterglow emission, but does not seem to be the norm. Table \ref{jets}
also shows that the uniform density inferred for those afterglows with
a good multiwavelength coverage is in the $0.05--10\, {\rm cm^{-3}}$
range, which is close to the particle density expected for a WR wind
at 0.1--1 pc. Thus, both the uniformity of burst ambient medium and
the density which we obtain indicate that the region where the 0.1--10
day afterglow emission is produced is the shocked WR wind (e.g. Wijers
2001). It remains to understand why the extent of the freely expanding
wind is so much smaller than expected.

The extent of the freely-streaming wind arising from the GRB
progenitor can be reduced from the extent predicted from our standard
model either by decreasing the strength of the wind, increasing the
density of the surrounding medium, or even increasing the pressure of
the surrounding medium (Chevalier et al. 2004; van Marle et al. 2005,
Fryer et al. 2006a). Higher densities and pressures in the surrounding
medium are more likely in regions of high-mass star formation, also
suggesting the progenitors of GRBs are in the high-mass end of
collapsing stars. Densities high enough to make a sufficiently small
wind bubble are generally higher than those allowed by radio
observations. Winds must be weakened substantially to make such a
small wind bubble.  Alternatively, removing the star from its original
wind bubble as happens in some merger models can produce appropriate
conditions.  Wolf-Rayet mass-loss rates depend sensitively on the
metallicity, and our above analyses assumed Wolf-Rayet mass-loss rates
comparable to what one would expect for solar metallicity stars.  To
induce black hole formation, the GRB progenitor scenarios in this
paper all predict that most GRBs are produced at low metallicities.
Single star models require modest winds to eject their envelopes, but
the winds of binary progenitors can be very weak.  The helium merger
progenitor can produce even weaker winds.

It is difficult to place strong constraints at this time from
observations of the surrounding medium.  The simple models used to
determine these quantities have many deficiencies so it is difficult
to be sure what the observations are truly telling us.  But the trend
indicates that an ideal progenitor scenario will tend to produce a
wind ejecta that is limited to the inner 1\,pc around the progenitor
star.  Further observations are required to provide a complete picture
on this observational constraint.

\subsection{Weak Supernovae associated with GRBs}

The co-spatial and concurrent observation of supernovae associated
with gamma-ray bursts provided the first convincing evidence that
gamma-ray bursts are produced in the collapse of massive stars and
propelled the collapsar model to the limelight as the leading model
for GRBs.  The luminosity of the supernova is roughly proportional to
the total amount of $^{56}$Ni produced in the explosion.  Recent
observations of two long-duration GRBs, produced in relatively typical
star-forming galaxies, have no observed supernova associated with the
gamma-ray burst (Fynbo et al. 2006, Berger et al. 2006).  Such weak
associated supernovae were predicted but, until this work,
indisputable claims of their existence could not be made (Fryer et al.
2006b).  Such observations have implications both for the mechanism
through which the black holes of GRBs are formed and the site of
nucleosynthesis for heavy elements such as $^{56}$Ni in GRBs.  The
observations imply that some GRBs produce very little ($\lesssim
0.07$\,M$_\odot$) $^{56}$Ni.  If this is true, this observation argues
for some GRBs to arise from systems whose black hole forms via
fallback, strongly constraining current models.

\subsection{Host Galaxy Morphology}

The morphological appearance of GRB host galaxies, particularly when
compared to the general population of star-forming galaxies (for
example in the HDF) may shed light on the environments and processes
that are conducive to the formation of GRB progenitors.  Hubble Space
Telescope observations are available for about 50 GRB hosts, and
recent visual and automated classification of these hosts lead to
several key results (Conselice et al. 2005; Wainwright et al. 2007).
First, we find that the radial light distribution of most GRB hosts is
exponential, as expected for disk galaxies (the median Sersic index is
about 1.1; Wainwright et al. 2007).  Second, the median effective
radius of the hosts is about 1.7 kpc, with a range of about 0.5-5 kpc.
Third, GRB host galaxies follow the size-luminosity trend observed in
other galaxy samples.  However, thanks to the relative faintness of
GRB hosts, and the ability to measure their redshifts independent of
the galaxy brightness, the GRB host sample extends the high redshift
size-luminosity relation by about 3 magnitudes (Wainwright et
al. 2007).

Most importantly, however, the overall morphological structure of the
host galaxies indicates an over-abundance of mergers/interactions
compared to star-forming galaxies in the HDF.  Overall, we find that
about 2/3 of all GRB host galaxies are morphologically disturbed.  In
the HDF, a similar fraction of galaxies appear to have an irregular
morphology, but only at $z>1$; at $z<1$, about 3/4 of all star-forming
galaxies have a regular morphology (Wainwright et al. 2007).  In
addition, the proportion of interacting galaxies in the field
increases with galaxy brightness (Conselice et al. 2003).  The fact
that the fraction of merging/interacting galaxies in the GRB sample is
independent of both redshift and galaxy brightness indicates that
these are regions of elevated star formation conducive to the
formation of GRB progenitors. As a corollary, it appears that GRBs are
less likely to occur in stable disk galaxies, and as a result GRB
hosts at low redshift are more likely to present a biased population
than at $z>1$.

\subsection{Distribution of Bursts with respect to Intensity}

The distribution of GRBs with respect to the light in a galaxy can
also place constraints on the progenitor.  Fruchter et al. (2006)
found that, whereas normal supernovae traced the light in a galaxy,
GRBs are actually more peaked toward the brightest regions in a
galaxy.  One of the more straightforward interpretations of this
observation is that GRBs arise from a more massive population of stars
than supernovae since the most massive stars are the most clustered
(see also our discussion in the last part of section 3.3).  It may
also imply that there is some feature of clustering that is required
to produce GRBs.  Exactly what this result teaches us awaits a much
more statistically significant set of observational data.

The strong constraint from this observation is that the progenitor 
system should arise from the most massive stars.  If true clustering 
can be proved, this might lead to new progenitor models.

\subsection{Gravitational Waves}  

Observations of gravitational waves can also help constrain the
progenitor by providing a direct probe of the angular momentum of the
collapsing star.  Instabilities in the accretion disk surrounding the
black hole and ringing in the forming black hole have both been
proposed as sources of gravitational waves (Fryer et al. 2002, see
Kobayashi \& Meszaros 2003 for a review).  Rockefeller et al. (2007)
found that, at least for some values of the angular momentum in the
collapsing stars, strong spiral instabilities can develop in the disk,
producing a gravitational wave signal that is over 10 times stronger
than the strongest rapidly-rotating normal supernova estimates
(consistent with the estimates from Fryer et al. 2002).  It is likely
that this signal will depend sensitively on the angular momentum of
the collapsing star and it can be used to constrain the rotation rates
of collapse progenitors.  However, if the signal is as weak as
Rockefeller et al. (2007) predict, it is unlikely we will have a
detection anytime soon.  van Putten (2005) has argued that the black
hole spin (with its enormous reservoir of energy) can couple to the
disk, producing strong instabilities that predict a signal that
should make many GRBs easily detectable by advanced LIGO.  If this
source is correct, the GW signal can easily be used to constrain
progenitor angular momenta.  Gravitational wave observations will first
determine which of these sources dominate the gravitational wave
signal from collapsars.  Once the source is determined, we can then
use gravitational wave observations to constrain the progenitor.

\section{Summary}

Since the first discovery of the optical counterpart to a GRB, there
has been a wealth of data pointing toward a massive star origin of
GRBs.  The collapsar engine, invoking the collapse of such massive
stars down to a black hole, has become the favored engine behind
long-duration gamma-ray bursts.  Although progenitors of this engine
have been studied for nearly a decade, the list of possible
progenitors is still large.   

We have reviewed many of the observations that may constrain the
nature of the progenitor.  Although the observations to date have
brought increased support for the massive star origin of gamma-ray
bursts, many of these observations are not strong enough to rule out
the progenitors.  But there are some strong statements that can be
made about the GRB progenitor.  The supernovae associated with GRBs
that are bright enough to be studied in detail are type Ic supernovae.
If this result is universal, any progenitor model must lose not only
its hydrogen envelope, but most of its helium envelope as well.  But
this sample is limited to 2-3 GRBs at the moment and it is also known
that some GRBs are not at all associated with bright supernovae.  We
also know that GRBs are even more clustered than their supernova
counterparts.  That is, they occur most often in the brightest parts
of bright galaxies.  This may just be an indication that these systems
only arise from the most massive stars, but better statistics may
argue that certain progenitors must take advantage of cluster
environments.  GRBs occur in environments with a range of
metallicities from 1/100th solar to solar.  The mean metallicity may
be as high as 1/3-1/2 solar.  This places strong constraints on
single-star models.  Finally, a potentially strong constraint on the
progenitor is the nature of the surrounding environment.  A number of
long-duration GRB progenitors require ejection of stellar material in
a strong wind for quite some time prior to collapse.  Current results
suggest that the circum-progenitor environment has a free-streaming
wind only out to 1\,pc; beyond that, the bulk of GRBs appear to have
constant density profiles.  If such a result can be solidified, it
places strong constraints on the GRB progenitor.

In this paper, we have reviewed many of the current collapsar
progenitors and their observational properties.  At this time, two
single-star progenitors exist: the classic single star scenario in
which the star loses its hydrogen envelope through winds, and the
mixing single star scenario in which the star is able to mix
sufficiently well to burn its hydrogen into helium.  Both of these
single star models have made strong predictions about the metallicity
requirements of the progenitor.  If taken at face value, both these
scenarios can be ruled out as sole progenitors of GRBs on metallicity
requirements alone.  These progenitors also do not fit the surrounding
environment extremely well, and also produce primarily He-rich (type
Ib) or even H-rich (type II) supernovae associated with GRBs.  These
single models clearly do not fit the existing data very well.

A large number of binary progenitors exist.  These models tend to fit
the metallicity constraints well.  Indeed, if we restrict ourselves to
the robust observational constraints, all of these progenitor
scenarios can match the existing data.  Some progenitor scenarios also
may fit the current data taken at face value (assuming all constraints
are robust) on the associated supernova, surrounding environment and
peaked clustering of GRBs.  But very few progenitor scenarios fit all
of these constraints without some tweaking.  There are obvious tweaks,
e.g. arguments why only a subset of these progenitors (that subset
that matches the strict interpretation of the data constraints) will
produce GRBs.  These scenarios will be differentiated as the
statistics in the current observations become stronger.  The current
ranking of the various progenitors when compared to the observations
is summarized in table~\ref{table:comparison}.

Studying the progenitors of type Ib/c supernovae may also provide some
insight into the progenitors of GRBs.  As with GRBs, when taken at
face value, current single-star stellar evolution models cannot
produce all normal type Ib/c supernovae.  In fact, the simulations by
Heger et al. (2003) argue that at solar metallicity, single stars
produce virtually no normal type Ib/c supernovae.  This argues
strongly that many Ib/c supernovae are produced in binaries.  Given
the high binary fraction of massive stars, this is to be expected and
it is unlikely that we will understand these supernovae well until
binary effects are added to stellar evolution codes.

With the existing robust constraints, it is unlikely that single-star
models can produce all GRBs.  As the data get better, the limitations
on single-star models will become more strict.  In addition, the data
have the potential to differentiate the currently proposed binary
progenitors.  As we focus in on a progenitor scenario, the metallicity
measurements of these GRBs may well teach us a lot about stellar
evolution.  But for this to work, we must not only obtain better
observational statistics; we have to refine our theoretical
understanding of these progenitors.  This requires a better physical
understanding of the uncertainties in stellar evolution and the
effects of binaries and the engine behind GRBs.  It is now within our
computational reach to understand mass loss, convection, and
magnetic fields in a rotating star, binary mass transfer and common
envelope evolution much better than our current parameterized models
allow.  This combined theoretical and observational work has the
potential, in the next two decades, to truly determine the GRB
generation.

\section{Definitions}

{\bf Case A,B,C Mass Transfer:} Close binary systems can undergo mass
transfer when one of the stars overfills its Roche radius.  The
``Case'' of this mass transfer is defined by the phase in the star's
life during which this mass transfer occurs: Case A - during main
sequence, Case B - after hydrogen burning but before helium ignition,
Case C - after helium ignition.

{\bf Collapsar:} The explosive engine that is powered by the collapse
of a massive star down to the black hole.  The energy is derived from
the potential energy released as a disk around this black hole
accretes onto the black hole.  This energy may be converted through
neutrinos and their subsequent annihilation or through magnetic fields
produced in the disk (Narayan et al. 1992 although see Fryer \&
M\'esz\'aros 2003).  All but one of the progenitor scenarios discussed
in this paper produce collapsars, that is, they produce a rapidly
spinning star that later collapses.  A similar progenitor scenario,
the ``helium-merger'' scenario (the merger of a compact object with a
helium star) is slightly different in that the compact remnant can be
formed long before the burst.  Nevertheless, it produces conditions
similar to those seen in the collapsar engine, and it is often lumped
into the collapsar category.

{\bf Common Envelope Evolution:} During mass-transfer, it is possible
that the matter overfilling the Roche radius accretes onto the
companion star faster than it can be incorporated into the companion
star or ejected from the system.  This material quickly forms an
atmosphere, or envelope, that surrounds both stars in the binary.
This ``common envelope'' phase leads to the rapid contraction of the
binary separation as tidal and viscous forces remove angular momentum
from the binary orbit.

{\bf Hypernova:} A super-energetic supernova explosion categorized by
those supernova-like explosions that exhibit higher energy and/or
larger asymmetries (more beaming) than normal supernova explosions.
GRBs are probably a subset of this class of explosion.

{\bf Supernova Types:} Supernova types are determined by observational
features.  Type I and II supernovae are distinguished by the presence
of hydrogen lines (type I have no hydrogen lines, type II have
hydrogen lines).  Type Ia are characterized by a deep silicon II
absorption line which is missing in type Ib/c supernovae.  Type Ib
supernovae have He I lines which are absent in Type Ic supernovae.  It
is believed that type Ib/c and II supernovae all arise from the
collapse of a massive star.

{\bf GRB Types:}  GRBs are classified by their duration and the 
hardness of their spectra.  The two primary classes are long ($>$1-3s) 
hard bursts and short ($<$1s), soft bursts.  There may be a third 
class of short, soft bursts (see Horvath et al. 2007 and references 
therein).

{\bf Acknowledgments} This project marks the culmination of many
discussions held at a workshop on the GRB/SN connection at the KITP
attended by nearly all of the authors and we are grateful for the
environment set up at the KITP that allowed these discussions and this
collaboration.  As such, this work is supported by the National
Science Foundation under Grant No. PHY99-07949.  It was also funded in
part under the auspices of the U.S. Dept. of Energy, and supported by
its contract W-7405-ENG-36 to Los Alamos National Laboratory, and by a
NASA grant SWIF03-0047.

{}

\newpage

\begin{deluxetable}{lcccc}
\tablewidth{0pt}
\tablecaption{Theoretical Predictions}
\tablehead{
  \colhead{Scenario}
& \colhead{Angular} 
& \colhead{Metallicity}
& \colhead{Surrounding}
& \colhead{Associated} \\

\colhead{}
& \colhead{Momentum}
& \colhead{Trend}
& \colhead{Environment}
& \colhead{Supernovae}
}
\startdata
Classic Single & Low? & Rate peaks $\sim 0.1 Z_\odot $ & High Wind & H-rich $\sim$ He-rich \\ 
Mixing Single & Good &  Z $< 0.1 Z_\odot$ & Low Wind & All He-rich \\ 
Classic Binary & Low?  & Rate$\uparrow$ $Z \downarrow$ & Tends to Low Wind & He-rich,He-poor \\
Tidal Binary & Good? & Rate$\uparrow$ $Z \downarrow$ & Tends to Low Wind & He-rich,He-poor \\
Brown Merger & Good? & Rate$\uparrow$ $Z \downarrow$ & Tends to Low Wind & He-rich,He-poor \\
Explosive Ejection & Good? & Rate$\uparrow$ $Z \downarrow$ & Shell within 1\,pc & He-poor \\
He-Merger & High & Rate$\uparrow$ $Z \downarrow$ & Tends to Low Wind & He-rich,He-poor \\
He-Case C & Good? & Rate$\uparrow$ $Z \downarrow$ & Tends to Low Wind & He-rich, more He-poor \\
Cluster & Good? & Rate$\uparrow$ $Z \downarrow$ & Tends to Low Wind? & He-rich? \\

\enddata
\label{table:progenitors}

\end{deluxetable}

\begin{deluxetable}{lll}
\tablewidth{0pt}
\tablecaption{Observational Constraints}
\tablehead{
  \colhead{Observation}
& \colhead{Strong Constraint} 
& \colhead{Trend}
}
\startdata
Rate & $0.001  < R_{\rm GRB}/R_{\rm Ib/c} < 0.1$ & $R_{\rm GRB} \approx 0.1 R_{\rm Ib/c}$ \\
Associated SN & Some are Ic & All are Ic \\
Metallicity & range: 0.01-1 solar &  mean $\sim$ 1/2-1/3 \\
Surrounding Environment & None Strong & Free-Streaming Wind limited to 1\,pc \\
Weak Supernovae & None Strong & Fallback BHs must occur \\
Host Morphology & None Strong & Interacting Galaxies (star formation/low metallicity?) \\
Distribution & $M_{\rm prog} \gtrsim 25M_\odot$ & Possibility Cluster Effects Important \\
Gravitational Waves & None Yet & Possibility to Constrain Angular Momentum \\

\enddata
\label{table:observation}

\end{deluxetable}

\begin{table*}
 \caption{ Reduced chi-square of the best fits obtained for 9 GRB
   afterglows (displaying optical light-curve breaks) with the Jet
   model (uniform outflow with sharp boundaries) and SO model
   (structured outflow with a power-law angular distribution of ejecta
   kinetic energy per solid angle), for a homogeneous medium (best fit
   density $n$ is uncertain by a factor 10) and a wind (density
   parameter $A_*$ uncertain by a factor 2-3) (from Panaitescu 2005) }
\label{jets}
\vspace*{5mm}
\centerline{ \begin{tabular}{lccccccccc}
  \hline
   GRB    & \multicolumn{2}{c}{Jet + n=const} & \multicolumn{2}{c}{Jet + wind}  
          & \multicolumn{2}{c}{SO + n=const}  & \multicolumn{2}{c}{SO + wind}  \\
          &  $\chi^2_\nu$ &  $n({\rm cm^{-3}})$ &  $\chi^2_\nu$ &  $A_*$      
          &  $\chi^2_\nu$ &  $n({\rm cm^{-3}})$ &  $\chi^2_\nu$ &  $A_*$           \\
  \hline
  980519  &  2.6  &  0.1   &   1.8  &  2.0   &   2.4  &  4   &   1.4  &  0.6   \\
  990123  &  2.0  &  0.8   &   2.3  &  0.2   &   1.8  &  1   &   2.2  &  0.1   \\
  990510  & 0.78  &  0.3   &   3.1  &  0.4   &   2.1  &  2   &   4.6  &  0.8   \\
  991216  &  2.0  &  0.04  &   1.8  &  0.2   &   1.2  & 0.7  &   1.2  &  0.4   \\
  000301c &  4.4  &  0.1   &   8.3  &  0.2   &   3.3  & 0.2  &   7.1  &  0.6   \\
  000926  &  2.2  &  20    &   3.5  &  1.8   &   2.2  &  3   &   2.8  &  0.5   \\
  010222  &  2.2  &  0.1   &   3.9  &  0.1   &   1.7  & 0.1  &   4.0  &   5    \\
  011211  &  4.7  &  1     &   8.7  &  0.6   &   2.3  &  1   &   4.7  &  0.7   \\
  020813  &  1.6  &  0.07  &   2.6  &  0.2   &   1.1  & 0.06 &   2.6  &  0.06  \\
  \hline 
\end{tabular} }
\end{table*}

\begin{deluxetable}{lcccccc}
\tablewidth{0pt}
\tablecaption{Theory vs. Observation\tablenotemark{a}}
\tablehead{
  \colhead{Progenitor}
& \colhead{Rate} 
& \colhead{Associated}
& \colhead{Metallicity}
& \colhead{Surrounding}
& \colhead{Host}
& \colhead{Distribution} \\

\colhead{}
& \colhead{}
& \colhead{Supernova}
& \colhead{}
& \colhead{Environment}
& \colhead{Morphology}
& \colhead{}
}
\startdata
Classic Single & T & s & s  & S & T & S \\
Mixing Single &  T & s & s  & S & T & S \\
Classic Binary & T & t & T & t & T & S \\
Tidal Binary & T & t & T & t & T & S \\
Brown Merger & T & t & T & t & T & S \\
Explosive Ejection & T & T & T & T & T & S \\
He-Merger & T & t & T & S & T & S \\
He-Case C & T & t & T & T & T & S \\
Cluster & T & t & T & t & T & T \\

\enddata 
\label{table:comparison}

\tablenotetext{a}{We rank the different scenarios by whether they pass
the Strong Constraint and the Trend (T) in the observations, the
Strong Constraint and the Trend with a modification or by using a
subset of the progenitor class (t), the strong constraint only (S), 
and the strong constraint only with modifications (s).  If the 
constraint has no strong constraint and the progenitor does not 
fit the trend, we use S.  These values are not set in stone.  Most 
rankings require much more detailed calculations to confirm.}

\end{deluxetable}
 

\begin{figure}
\plotone{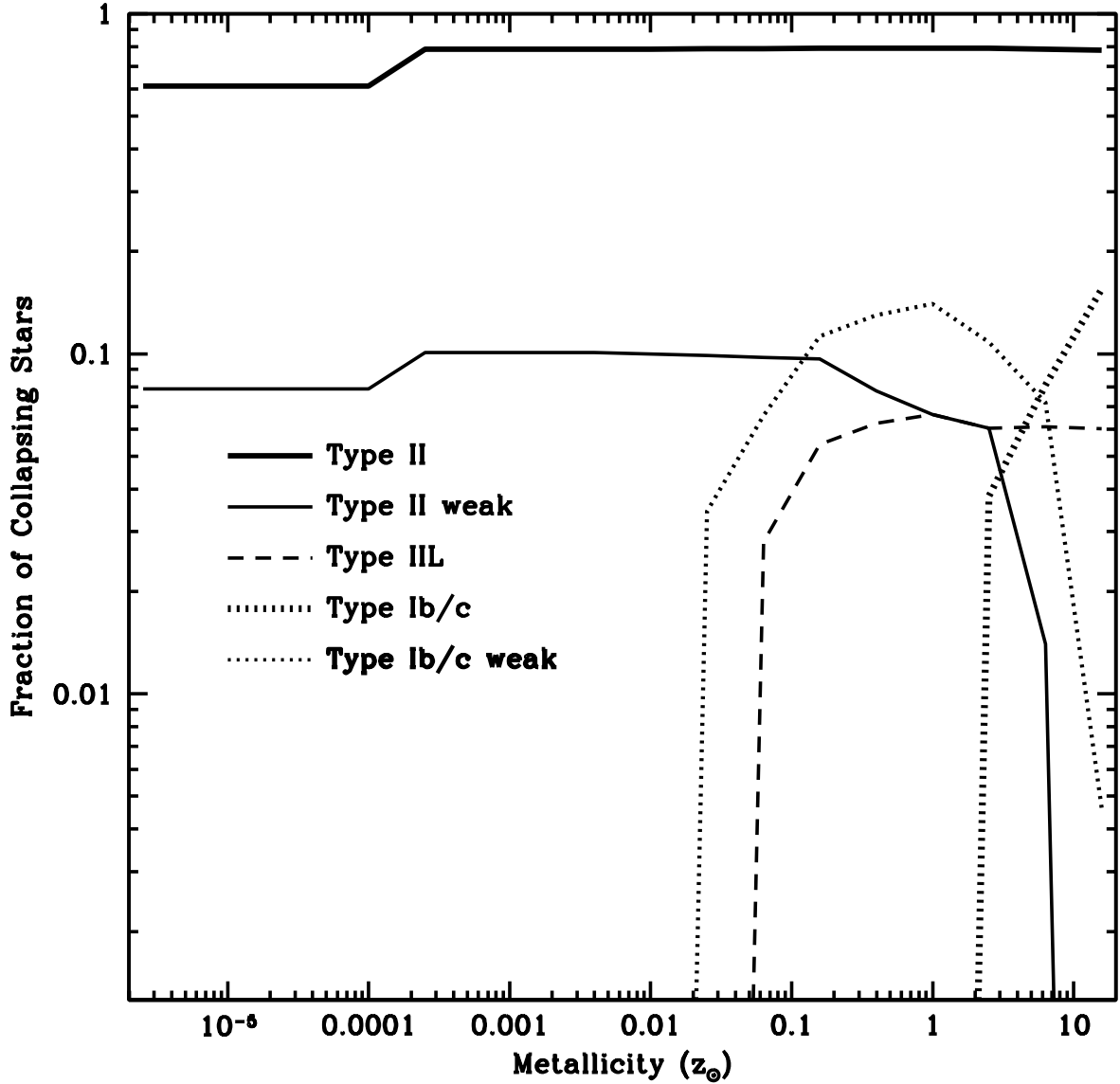}
\caption{Single-star supernova rate as a fraction of total number of
collapsing stars as a function of metallicity calculated using the
stellar models from Heger et al.(2003).  We consider 3 classes of type
II supernovae: normal type II supernovae (Type II), Type II supernovae
with weak supernova explosions (Type II weak), Type II supernovae that
have lost most of their hydrogen envelope (Type IIL) and two classes
of type Ib/c supernovae: normal Ib/c supernovae and weak Ib/c
supernovae.  If single stars dominate the Ib/c rate, these models
predict only weak type Ib/c supernovae below solar metallicity.}
\label{fig:singleSN}
\end{figure}
\clearpage

\begin{figure}
\plotone{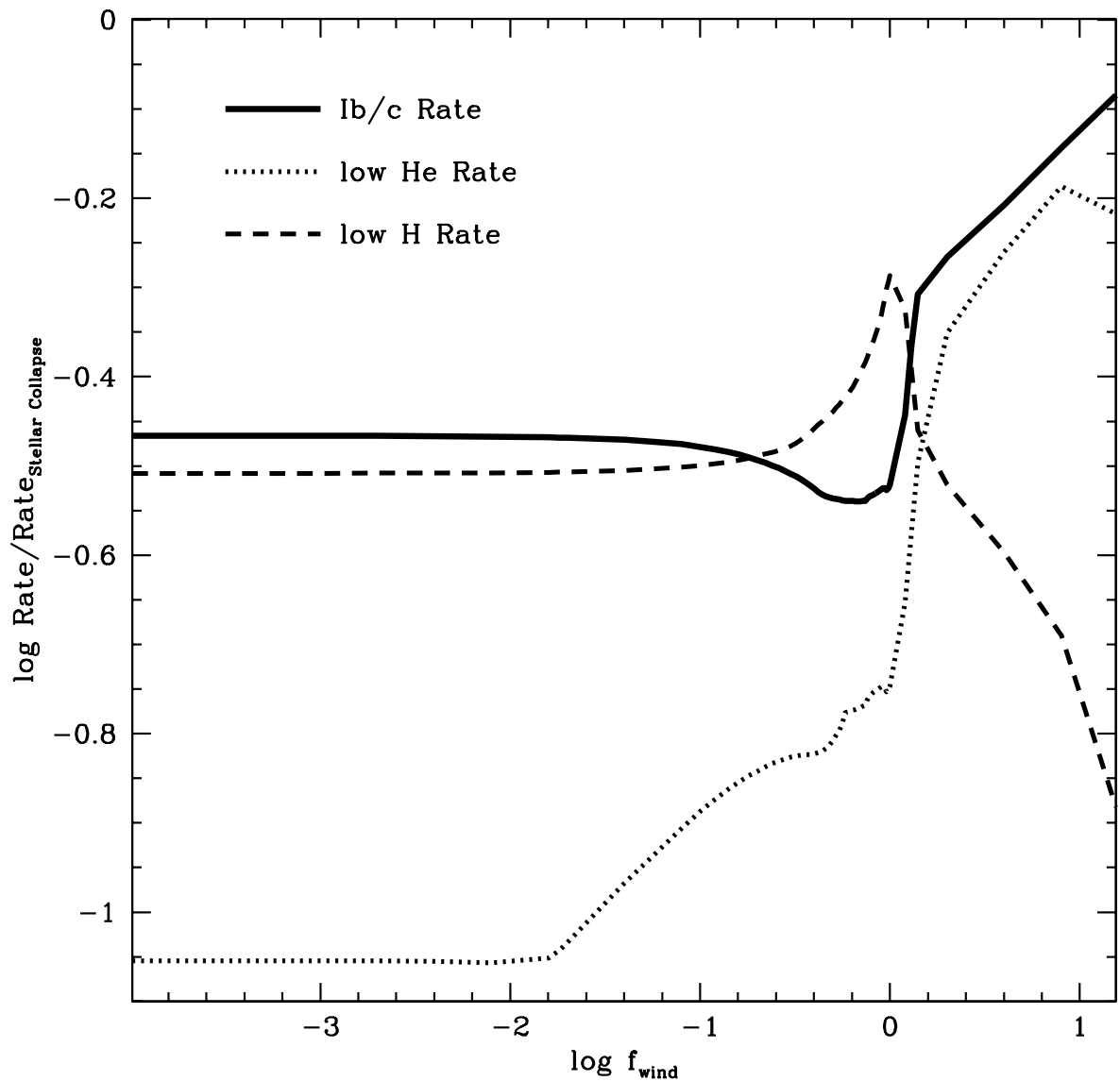}
\caption{Binary-star supernova rate as a function of the mass loss
parameter (see Fryer et al. 1998 for details).  We consider three
types of supernovae: all type Ib/c supernovae, type Ib/c supernovae
that have lost more than 1\,M$_\odot$ of helium (not necessarily type
Ic supernovae, but on their way to becoming Ic supernovae), and type
II supernovae that have lost 2/3 of their hydrogen (peculiar type II
or type IIL supernovae) but still retain some hydrogen envelope to be
type II supernovae.  As the mass-loss parameter increases, the rate of
Ib/c and low-helium supernovae increases while the rate of low
hydrogen (and all type II supernovae for that matter) decreases.  This
mass-loss parameter can be seen as a parameter for the metallicity.
Although the rate of Ib/c supernovae is higher at higher mass-loss
rates (higher metallicities), it drops less than a factor of 5 when
varying the mass-loss parameter from very high mass-loss to
essentially no mass-loss rates.} 
\label{fig:binarySN}
\end{figure}
\clearpage

\begin{figure}
\plotone{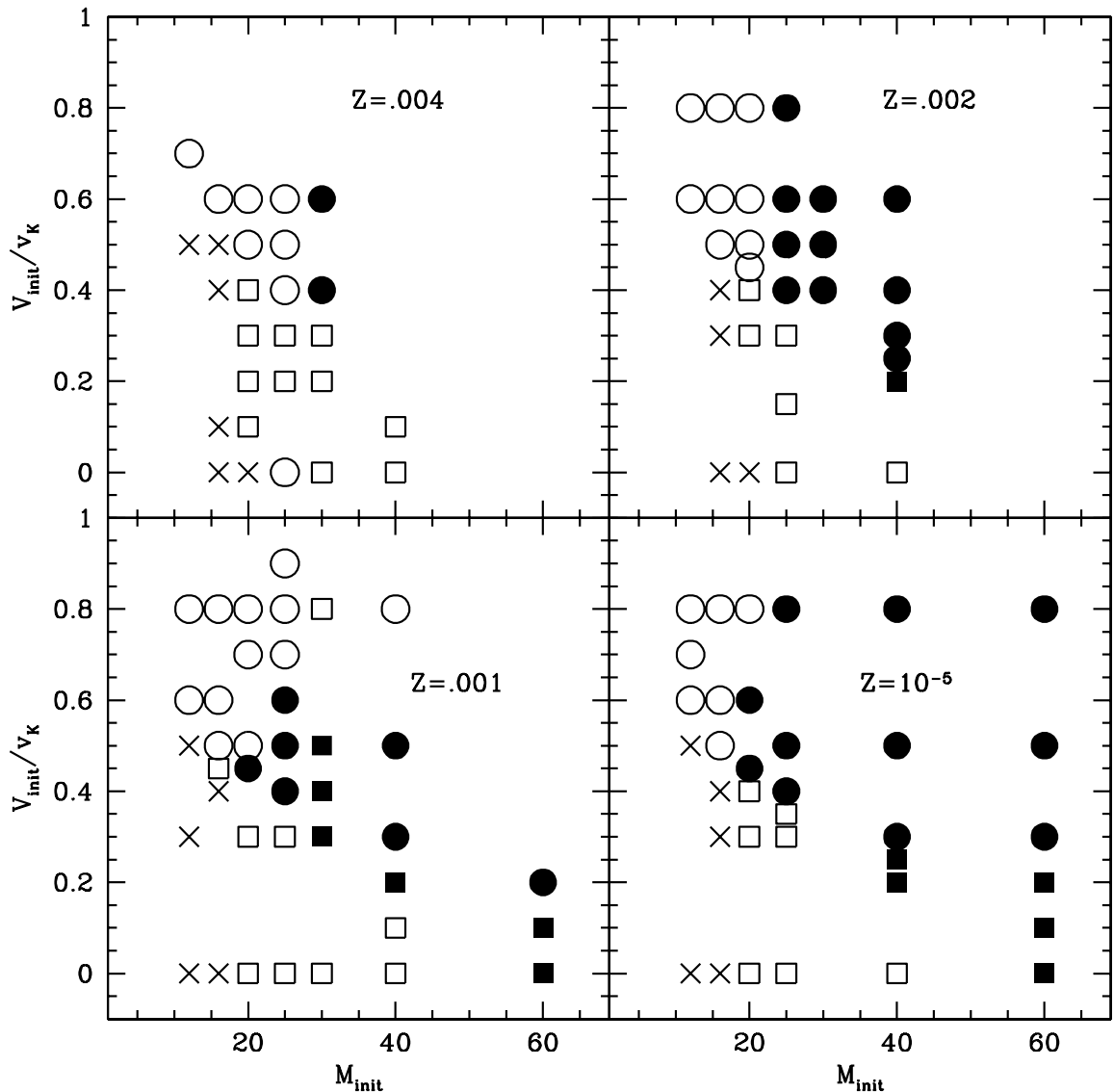}
\caption{Fate of massive single stars as a function of initial mass
and spin period.  The squares correspond to type II (H rich)
collapses, the circles correspond to type Ib/c (H-deficient)
collapses.  Filled square/circles correspond to direct collapse
objects, open square/circles correspond to weak supernova explosions,
and the crosses correspond to normal supernova explosions.  There are
no normal Ib/c supernovae produced with this particular grid of stars
from Yoon et al. (2006).  These stellar models ended at core C/O
burning, so we do not have a collapsed core to examine to determine
its true fate.  Instead we use the C/O core mass, using the Fryer
(1999) analysis and comparing the cores of those collapsing C/O cores
to the C/O cores presented by Yoon et al.(2006).  For normal stars,
Fryer (1999) predicts stars with low mass-loss and initial masses
above roughly 20\,M$_\odot$ will produce weak supernova explosions and
black holes, stars above 45\,M$_\odot$ do not produce supernovae
explosions at all (although both these types of objects may produce
GRBs and their associated supernovae).  The size of the C/O core
varies from simulation to simulation.  The Limongi \& Chieffi(2006) C/O
cores tend to be 20\% lower than their Woosley et al.(2002)
counterparts.  The Yoon et al.(2006) cores tend to be 30\% smaller, 
but this is, in part, due to the fact that no post-C/O core ignition 
shell burning contributes to the C/O core mass.  We chose a 4\,M$_\odot$ 
C/O core mass for the dividing line between strong and weak supernova 
explosions and a 13\,M$_\odot$ dividing line between weak and no supernova 
explosions, consistent with the Fryer (1999) analysis.  30\% variations 
in this mass limit will not vary the results significantly.}
\label{fig:yl}
\end{figure}
\clearpage

\begin{figure}
\plotone{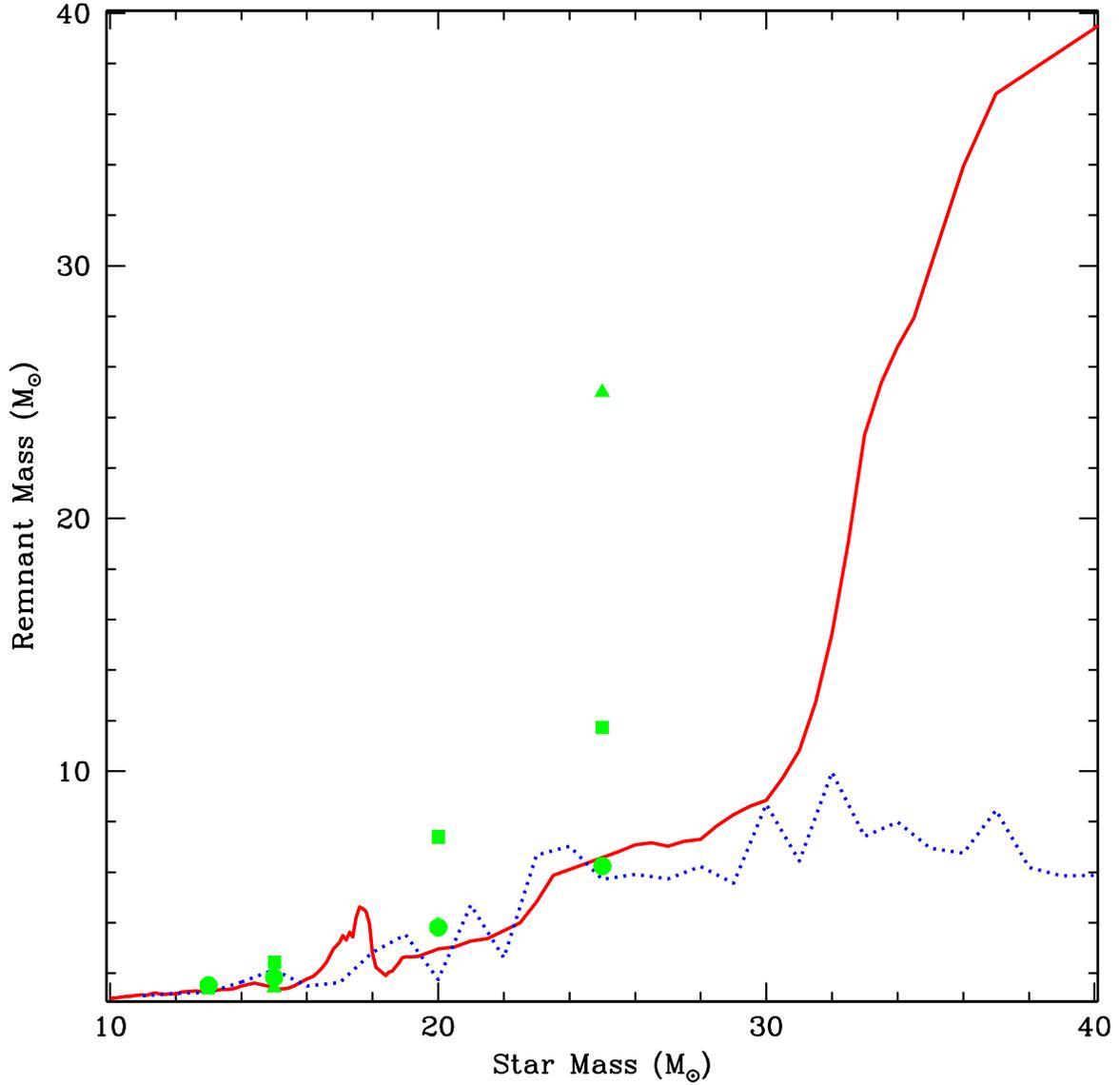}
\caption{Remnant mass using the Fryer (2006) analysis versus initial
star mass for both the Limongi \& Chieffi (2006) and the Woosley et
al. (2002) stellar progenitors.  The lines are derived from the
Woosley et al. (2002) progenitors: dotted line refers to solar
metallicity, solid line refers to very low metallicity.  The points
are derived from the Limongi \& Chieffi (2006) models: circle - solar,
square - 0.2 solar, triangle - zero, metallicities.  Around
20\,M$_\odot$, the fate of the stars depend sensitively upon the 
stellar evolution code used.  However, it is clear that around 
20\,M$_\odot$ is roughly the dividing line between neutron star 
and black hole formation.}
\label{fig:remnant}
\end{figure}
\clearpage

\begin{figure}
\plotone{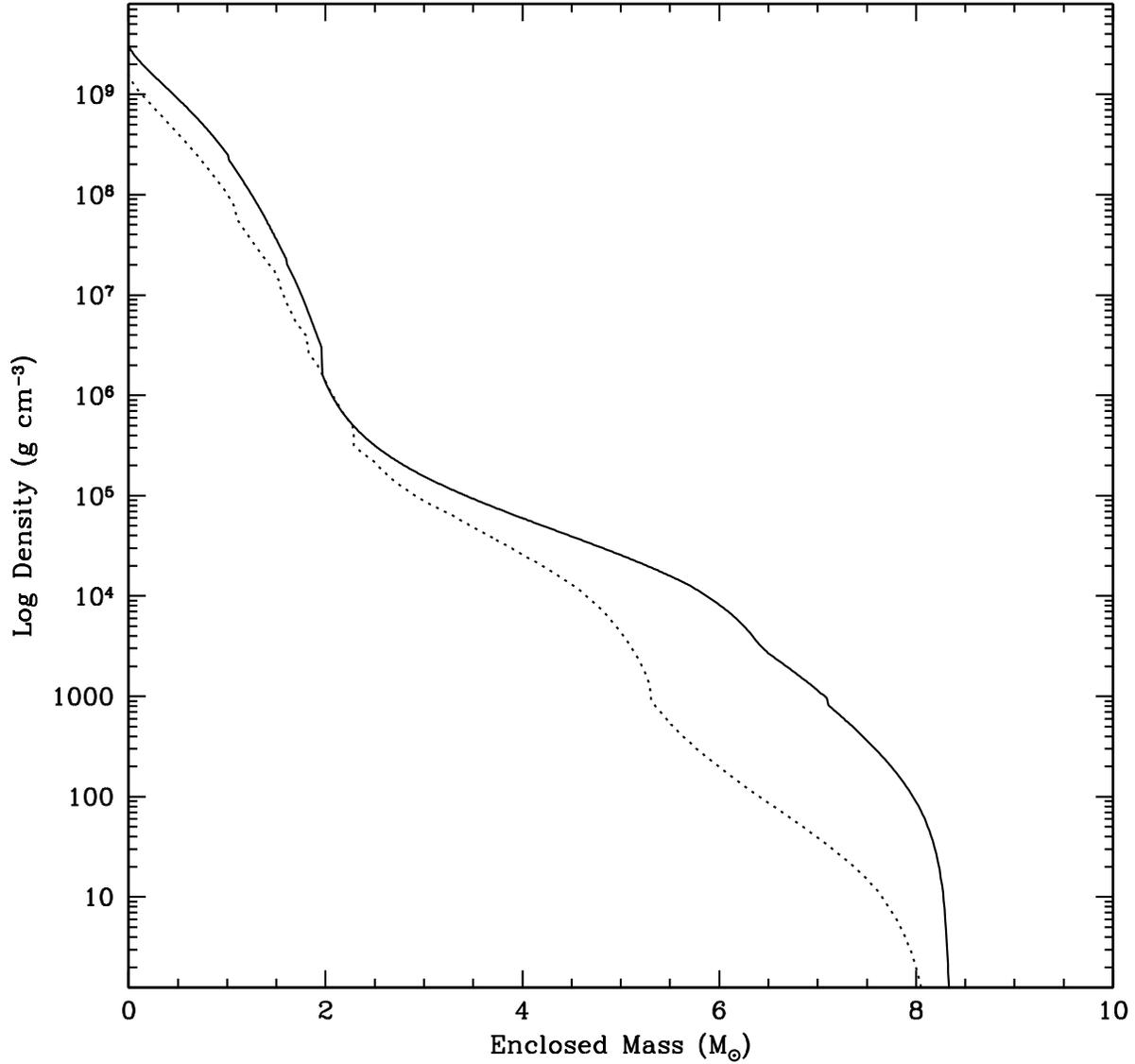}
\caption{Density profiles of the Limongi \& Chieffi (2006) and 
Woosley et al. (2002) 25\,M$_\odot$ stars at collapse.  The 
differences in the inner 1\,M$_\odot$ can be explained by the 
fact that the models are at different stages in the collapse.  
A difference of a fraction of a second can cause the difference 
in densities in this inner region.  But the differences beyond 
this inner core (beyond 2\,M$_\odot$) can only be explained 
by uncertainties in the stellar evolution models.}
\label{fig:density}
\end{figure}
\clearpage

\begin{figure}
\plotone{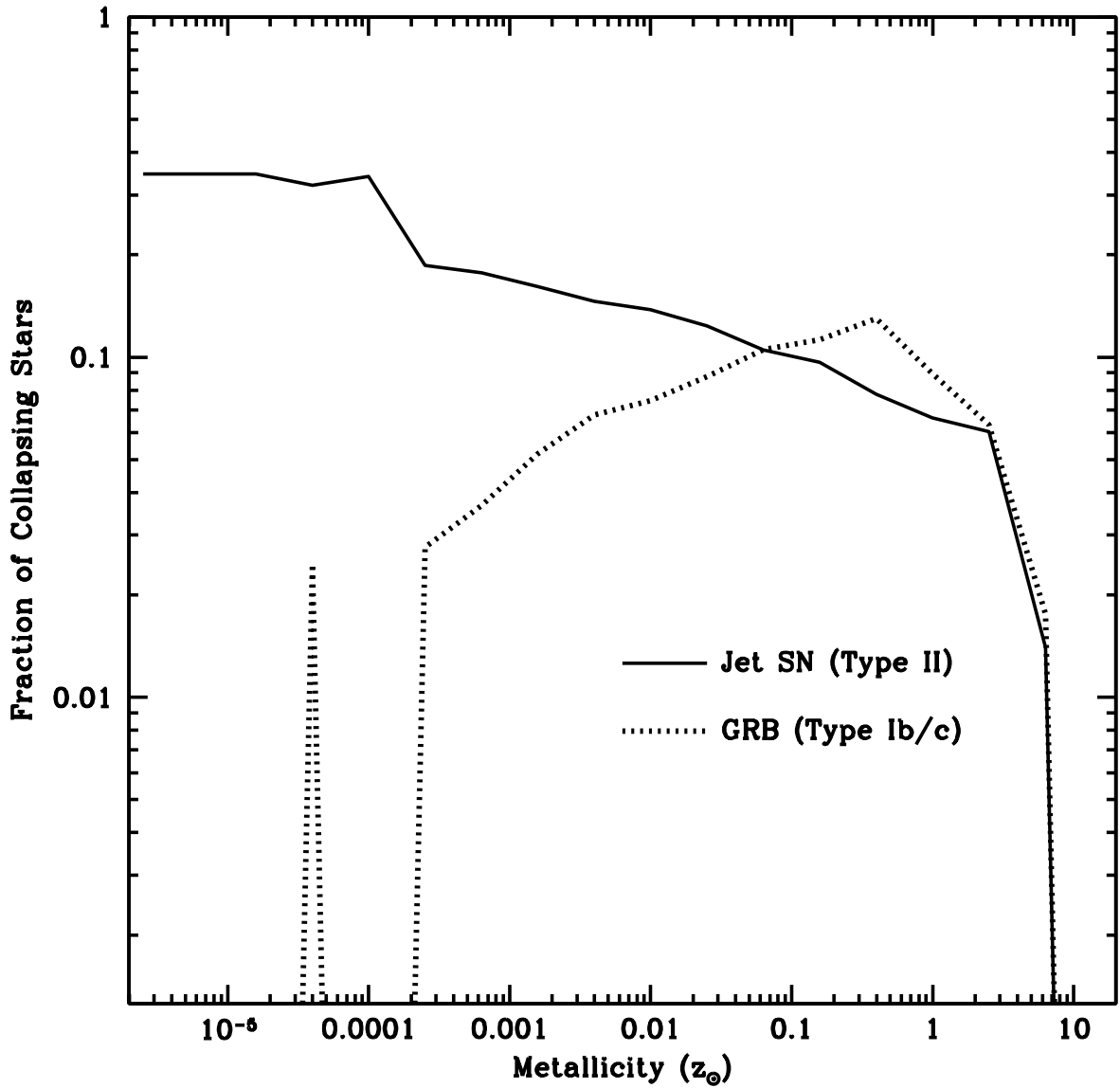}
\caption{GRB rate as a function of collapsing stars for single stars
as a function of metallicity using the Heger et al.(2003) models.  The
solid line shows those possible hypernovae with hydrogen envelopes
(termed ``jet'' supernovae by Heger et al. 2003).  The dotted line
shows systems that lose their hydrogen envelopes and hence {\it could}
be GRB progenitors.  These numbers must be multiplied by a factor
indicating what fraction of these stars actually retain enough angular
momentum to make black hole accretion disks (could be 0).  To make
single models work, we must somehow explain why we don't observe jet
supernovae (currently a matter of debate in stellar evolution
theory).}
\label{fig:GRB}
\end{figure}
\clearpage

\end{document}